\documentclass[paper, 12pt, letterpaper]{JHEP} \input{epsf.tex}
\usepackage{graphics} \usepackage{epsfig}
\usepackage{amsmath}\usepackage{amssymb}

 \def\C{{\mathbb C}} \def\R{{\mathbb R}} \def\Z{{\mathbb
    Z}}   \def\P{{\mathbb P}}

 \def\Re {\bb{R}}  
            
 \def\Vol{\operatorname{Vol}}
 \def\tr{\operatorname{tr}}

  \def\Im{{\rm Im\,}}
 \def\Re{{\rm Re\,}} \def\tr{{\rm tr\, }}
\def\Tr{{\rm Tr\, }} \def\d{{\rm d }}

\def\id{\protect{{1 \kern-.28em {\rm l}}}}

\newcommand{\be}{\begin{equation}} \newcommand{\ee}{\end{equation}}
\newcommand{\bea}{\begin{eqnarray}} \newcommand{\eea}{\end{eqnarray}}
\newcommand{\beann}{\begin{eqnarray*}}
  \newcommand{\eeann}{\end{eqnarray*}}
\newcommand{\bfig}{\begin{figure}} \newcommand{\efig}{\end{figure}}
\newcommand{\nn}{\nonumber}
\newcommand{\ba}{\begin{array}}\newcommand{\ea}{\end{array}}

\newtheorem{Proposition}{Proposition}[section]

\newtheorem{Theorem}{Theorem}[section]
\newtheorem{Lemma}{Lemma}[section]
\newtheorem{Corrolary}{Corrolary}[section]

\newcommand{\bp}{\begin{Proposition}}
  \newcommand{\ep}{\end{Proposition}}
\newcommand{\bt}{\begin{Theorem}} \newcommand{\et}{\end{Theorem}}
\newcommand{\bl}{\begin{Lemma}} \newcommand{\el}{\end{Lemma}}
\newcommand{\bc}{\begin{Corrolary}} \newcommand{\ec}{\end{Corrolary}}

\newcommand{\Mat}{\operatorname{Mat}} \newcommand{\eps}{\varepsilon}
\newcommand{\erf}[1]{(\ref{#1})} \newcommand{\imp}{\operatorname{Im}}

\newcommand{\la}{\lambda} \newcommand{\om}{\omega}
\newcommand{\rep}{\operatorname{Re}}
\newcommand{\void}[1]{}

  \def\om{\omega}
   \def\ep{\eps}

\title{Constructing Gauge Theory Geometries from Matrix Models}

\author{A. Klemm, K. Landsteiner, C. I. Lazaroiu, I. Runkel\\
  Humboldt Universit\"at zu Berlin\\
  Insitut f\"ur Physik\\
  Invalidenstrasse 110, Berlin\\ Germany\\
  aklemm, calin, ingo, landsteiner@physik.hu-berlin.de}

\abstract{We use the matrix model --- gauge theory correspondence of
  Dijkgraaf and Vafa in order to construct the geometry encoding the
  exact gaugino condensate superpotential for the  ${\cal N}=1$ 
  $U(N)$ gauge theory with adjoint and
  symmetric or anti-symmetric matter, broken by a tree level 
  superpotential to a product subgroup involving $U(N_i)$ and $SO(N_i)$ or
  $Sp(\frac{N_i}{2})$ factors. The relevant geometry is encoded by a
  non-hyperelliptic Riemann surface, which we extract from
  the exact loop equations. We also show that $O(1/N)$ corrections
  can be extracted from a logarithmic deformation of this surface.
  The loop equations contain explicitly subleading terms
  of order $1/N$, which encode information of string theory on an
  orientifolded local quiver geometry. }

\preprint{HU-EP-03-01}

\begin{document}

\pagebreak

\vskip .6in

\section{Introduction}

Holomorphic terms in the effective action of supersymmetric gauge
theories give much desired exact information about their
non-perturbative dynamics. Based on geometric engineering through
string compactifications, Dijkgraaf and Vafa \cite{DV1,DV2,DV3}
conjectured that the exact holomorphic superpotential of certain
effective ${\cal N}=1$ gauge theories\footnote{ Some of these results
  carry over to ${\cal N}=0$ orbifolds \cite{DNV} of ${\cal N}=1$
  gauge theories.}  is calculated by the planar diagrams of a matrix
model.  In the simplest case, an ${\cal N}=2$ $U(N)$ gauge theory is
broken by a tree-level superpotential $W(\phi)=\sum_{i=1}^n
\frac{g_k}{k+1} {\rm Tr} \phi^{k+1}$ to an ${\cal N}=1$ theory with
gauge group $G=\prod_{i=1}^r U(N_i)$. Here $\phi$ is an adjoint chiral
superfield, and one assumes that all zeros of $W$ are simple and
that $\sum_{i=1}^r N_i=N$.  An effective theory emerges by integrating
out the massive chiral fields.  Its effective low energy Lagrangian
can be expressed in terms of the gaugino bilinear fields $S_i
=-\frac{1}{32 \pi^2} {\rm Tr}\, W_{\alpha\, i} W^{\alpha\, i}$
where $i$ indicates the factors of the unbroken gauge group.  The
claim is that the effective action as a function of these fields is
calculated exactly by the leading terms in the $1/N$ expansion
of a matrix model whose matrix potential is given by $W$.

Technically the field theory results for the case mentioned above are
not new. They were obtained by geometric engineering in
\cite{Cachazo:2001jy} and partly by factorization of the ${\cal N}=2$
Seiberg-Witten curve \cite{Cachazo:2001jy, Cachazo:2002pr}. This
allows for explicit checks \cite{DGKV} and leads to a
particular interpretation of the matrix model. The relevance of matrix
model structures, such as the planar loop equations, can also be
understood more conceptually using supergraph techniques
\cite{Dijkgraaf:2002xd} or anomalies \cite{gorsky, Cachazo:2002ry}.
Further developments can be found in \cite{phases, seiberg, radu, binor}.

In this paper, we shall add matter in the symmetric or antisymmetric
representation of the $U(N)$ gauge group and consider an additional
term ${\rm Tr}\ \bar Q \phi Q$ in the tree-level superpotential
\footnote{Symmetric and antisymmetric representations are interesting
  because they are building blocks of chiral supersymmetric gauge
  theories. In particular many models with dynamical supersymmetry
  breaking are based on antisymmetric representations, see e.g.
  \cite{ADS,MV,Leigh:1997sj}.}. This leads to a richer vacuum
structure, which has branches where additional $SO(N_i)$ and
$Sp(N_i/2)$ gauge factors appear. The conjecture of \cite{DV1} then
gives us a rough recipe for how to obtain a dual matrix model.
However, supersymmetric vacua are described by a symplectic quotient,
which is equivalent with a holomorphic quotient of the space of
$F$-flat configurations through the action of the 
complexified gauge group.  Thus
the gauge-theory analysis of the vacuum structure leads to solutions
in the space of complex matrices modulo complexified gauge
transformations.  On the other hand, Hermitian matrix models have
hermiticity constraints on their matrix variables. To make the matrix
model useful for the gauge theory analysis, an appropriate constraint
should be imposed in a way which allows one to recover the correct
vacuum structure of the gauge theory. As an example, consider the 
one-matrix model associated with the $U(N)$ gauge theory broken by a cubic
superpotential $W(\phi)$.  Then the field theory has two classical vacua at
$W'(\phi)=0$, which, for a Hermitian matrix $\phi$, would correspond
to a local maximum and a local minimum for its eigenvalues. Perturbation
theory around the local maximum would make no sense in the Hermitian
model.  In this case, the relevant information was extracted in
\cite{KMT} and \cite{Detal}, where it was shown that it matches the
gauge theory instanton expansion relevant for both vacua.  Using the
BRST formalism, the residual gauge symmetry $U(N)/(U(N_1)\times
U(N_2))$ acting on the vacuum leads to an expansion scheme in terms of
a two-matrix model interacting through the ghosts \cite{DGKV}. To avoid
convergence problems, one must reinterpret the expansion around the extrema in
terms of a Hermitian $N_1\times N_1$ matrix interacting with an {\em
anti-Hermitian} $N_2\times N_2$ matrix \cite{KMT}. The resulting perturbative
expansion then recovers the leading $F$-term gravitational corrections to the
gauge theory, which in the matrix model arise at order $O(1/N^2)$ \cite{KMT,
Detal}.  A more systematic analysis of such issues requires a holomorphic
construction of the relevant matrix model \cite{calin}. Namely it leads to a
so-called `holomorphic matrix model', whose relevance was already pointed out
in \cite{DV1}.  As shown in \cite{calin}, the procedure employed in \cite{KMT}
is indeed perfectly justified in the holomorphic matrix model.  In the present
paper, we shall encounter similar problems even before choosing a
vacuum. Accordingly, we must carefully choose a `real section' through the
space of complex matrixes in order to make all integrals converge. Moreover,
we shall need a complex regularization of the matrix model in order to avoid
introducing spurious constraints on its filling fractions.  This is very
similar to what happens in the case of holomorphic $A_2$ models \cite{calin},
where use of a complex regularization is crucial in order to avoid similar
problems.

In section two we introduce the gauge theory model and analyze its
classical vacuum structure. Using threshold matching, we also derive
the leading logarithmic terms in the effective superpotential.

In section three we define the corresponding matrix model, whose
action can be taken to be the superpotential of the gauge
theory.  As mentioned above, it will be crucial to choose a suitable
replacement for the hermiticity constraint. Contrary to the one-matrix
model, we cannot impose hermiticity of the matrix $M$ which
corresponds to the adjoint chiral multiplet $\phi$.  Indeed, this
requirement would prevent the matrix model from probing the
complete vacuum structure of the dual gauge theory.  As explained in
\cite{calin}, this is not a particular property of our model, but can
be traced back to similar issues encountered in $ADE$ matrix models.
Using a holomorphic definition and a 
`complex' regularization of our model, we shall extract {\em two} exact loop
equations (Ward identities) which are cubic and quadratic in the 
resolvent. A new
feature of these relations is that they contain explicitly terms of
order $O(1/N)$ and $O(1/N^2)$.  Taking the large $N$ limit of these
constraints allows us to extract a non-hyperelliptic Riemann surface
which governs the planar solutions of the model, as well as the
effective superpotential of the dual field theory.  The dual string
geometry is expected to be given by an orientifold of the non-compact 
Calabi-Yau background dual to the $A_2$
quiver theory \footnote{Factorization of the ${\cal N}=2$ curve cannot
  give hints, since it leads to a completely different branch of the
  moduli space.}.  We therefore also give the explicit derivation of
the full loop equations for the matrix model based on the $A_2$
quiver, and relate it to the $\mathbb{Z}_2$ orbifold and orientifolds
thereof.  However, there are subtleties in the orientifolding, in the
large $N$ transition and in the precise definition of the $B$-model
cycles.  We therefore prefer to use matrix model techniques 
in order to extract the Riemann surface, which
should already encode all relevant information
about the holomorphic terms in the ${\cal N}=1$ effective action.
This surface is a triple cover of the complex plane, which
cannot generally be written as a hyperelliptic curve. Since we have
little guidance from a string geometry, we must intrinsically understand
this surface, which in the geometric engineering approach 
would be obtained by integrating out two
directions of a dual  Calabi-Yau geometry \cite{Cachazo:2001jy}. 
In particular, one must take into account issues of non-compactness 
and relations between periods, which are important for a 
proper count of this curve's parameters. 

Similar to the case of $SO(N)$ and $Sp(\frac{N}{2})$ groups \cite{Ita:2002kx,
Ashok:2002bi, so1, so2, so3, so4, so5, so6, so7}, the 't Hooft expansion leads
to unoriented ribbon graphs and the large $N$ expansion of the matrix model is
spaced by $N$ and not $N^2$.  Thus the explicit $1/N$ terms entering our loop
equations have field-theoretic relevance.  To properly analyze such effects,
one must implement the filling fraction constraints $N_i=$const of \cite{DV1}
in the finite $N$ model, which will be done in Section 4 by introducing
chemical potentials and performing a Legendre transform to extract a
microcanonical generating function.  This gives a direct derivation of a set
of special geometry relations, and provides their finite $N$ generalization.
It also naturally leads to a finite $N$ version of Whitham-type constraints.

In Section five we show how $O(1/N)$ contributions can be
computed from a modified Riemann surface, which is obtained by
performing a logarithmic deformation of the matrix model. 
We show that the $O(1/N)$ term of the microcanonical
generating function can be computed by differentiating the leading
($N=\infty$) contribution with respect to the coupling constant of
such a logarithmic deformation.

In Section 6, we check agreement between the matrix model and the
field theory. For this, we use the matrix model to compute the leading
(Veneziano-Yankielowicz) contribution to the effective superpotential for
different branches of the moduli space. This is done by performing
BRST gauge-fixing in the manner of \cite{DGKV, KMT} and integrating
out the quadratic terms in the action.
The comparison to the $O(N)$ model and the $A_2$-quiver is summarized
in the Appendix.

\section{A first view of field theory properties}
\label{fieldtheoryI}

In this section we study $U(N)$ ${\cal N}=1$ supersymmetric gauge
theory with matter in the symmetric or in the antisymmetric
representation. The starting point is an ${\cal N}=2$ $U(N)$ gauge
theory with matter in the symmetric or antisymmetric representation
$(Q^T,\bar Q^T)=s(Q,\bar Q)$. We choose a tree-level superpotential
\be \mathcal{W} = \tr ( W(\phi) + \bar{Q} \phi Q )\, ,
\label{treelevelsupo}
\end{equation}
which softly breaks the ${\cal N}=2$ supersymmetry to ${\cal N}=1$. In
many of our considerations below it helps to view the theory as coming
from a $\Z_2$-orientifold of the $A_2$ quiver theory.  Orientifolds of
supersymmetric $A_2$ quiver gauge theories have been constructed
before with the help of Hanany-Witten type brane configurations in
Type IIA string theory \cite{Landsteiner:1997ei,Landsteiner:1998pb}.
These models were further analyzed in
\cite{Krichever:1999zu,Krichever:2001ji,Naculich:1998rh,Ennes:1998gh}.
Whereas an $A_2$ quiver gauge theory contains two independent unitary
gauge groups and matter that transforms under the bifundamental
representation, the orientifolded model identifies the two gauge
groups. In addition the orientifold modifies how the matter content
transforms.  In the $A_2$ model the chiral superfield $Q$ transforms
as $U_1 Q U_2^\dagger$ where $U_1$ and $U_2$ are the gauge
transformation of the two independent gauge groups.  In the
orientifolded model this changes to $Q\rightarrow U Q U^T$.  Therefore
$Q$ transforms in a two-index tensor representation of $U(N)$ and the
two possibilities of symmetric or antisymmetric representation
correspond to the two choices of orientifolds.

In the following we describe the vacuum and its phase structure of the
emerging ${\cal N}=1$ theory and calculate in various phases the
Veneziano-Yankielowicz potential from threshold matching.

\subsection{The classical moduli space}
We are going to classify the possible constant solutions of the
classical field equations.  Throughout the analysis we also assume
that ${\cal N}=1$ Fayet-Iliopoulos terms are absent. As is well known
in this case the space of solutions of the field equations can be
obtained by minimizing the superpotential $W$ and dividing by the
complexified gauge group. The field equations are
\begin{eqnarray}
Q \bar{Q} + W'(\phi) &=&0\,,\label{eomphi}\\
\phi Q +  Q \phi^T &=&0\,,\label{eomq}\\
\bar{Q} \phi + \phi^T\bar{Q}&=&0\,.\label{eomqb}
\end{eqnarray}
For quiver gauge theories similar field equations arise and the
classical moduli space has been analyzed in \cite{Cachazo:2001gh}. We
will adapt the methods there to the case at hand.

Note first that from (\ref{eomq}), (\ref{eomqb}) it follows that
\begin{eqnarray}
\phi^n Q = Q (-\phi^T)^n \,,\nonumber\\
\bar{Q} \phi^n = (-\phi^T)^n \bar{Q}\,,
\end{eqnarray}
We multiply (\ref{eomphi}) from the left with $\bar{Q}$, from the
right with $Q$ and commute $Q$ through $W'$ with the help of the
previous relation.  This gives $\bar{Q}Q [ \bar{Q} Q + W(-\phi^T)]=0$.
Now we transpose (\ref{eomphi}) and use it to eliminate $\bar{Q}Q$ and
we obtain $ -W'(\phi^T)[-W'(\phi^T)+W'(-\phi^T)]=0$. Finally we
transpose this last equation and arrive at
\begin{equation}\label{eomi}
[W'(\phi)-W'(-\phi)]W'(\phi)=0
\end{equation}
Equation (\ref{eomi}) will be solved for $\phi$ being diagonal $\phi =
\mathrm{diag}(a_1.{\bf{1}}_{N_1},\cdots,a_n.{\bf{1}}_{N_n})$ and
$\sum_{i=1}^n N_i =N$.  The entries $a_i$ have to be the roots of one
of the two equations
\begin{eqnarray}
W'(x)-W'(-x)&=&0\label{rootsi}\,,\\
W'(x)&=&0\,\label{rootsii}\,.
\end{eqnarray}
The vev of $\phi$ breaks the gauge group according to $U(N)\rightarrow
\prod_{i=1}^n U(N_i)$. The field $Q$ decomposes into $Q_{ij}$ with
$Q_{ij}$ transforming as a bifundamental $(N_i,\bar{N}_j)$ under
$U(N_i)\otimes U(N_j)$ if $i\neq j$ and as symmetric (antisymmetric)
if $i=j$. An analogous statement holds for $\bar Q$. {}From
(\ref{eomq}) it follows that
\begin{equation}
(a_i + a_j)Q_{ij}=0
\end{equation}
and thus $Q_{ij}=0$ unless $a_i=-a_j$ and $N_i=N_j$.  Such pairs of
solutions are indeed generated by the roots of (\ref{rootsi}).
Another special root of (\ref{rootsi}) is $x=0$.  Let us study now in
more detail the different solutions.

\subsubsection{Pair of solutions $a_i=-a_j=b\neq 0$ with $W'(b)=W'(-b)$}
In the relevant subspace $Q$ and $\bar{Q}$ have to be of the form
\begin{equation}
Q = \left(\begin{array}{cc} 0&q\\
s  q^T &0 \end{array}\right)\qquad
\bar Q = \left(\begin{array}{cc} 0&\bar q \\
s  \bar{q}^T &0 \end{array}\right)\,.
\end{equation}

A gauge transformation acts on $Q$ as
\begin{equation}
Q\rightarrow \left(\begin{array}{cc} 0&U_1 q U_2^T\\ s  U_2q^T U_1^T &0
\end{array} \right)\qquad
\end{equation}
Since $U_1$ and $U_2$ are independent $GL(N_i,\mathbb{C})$ matrices we
can bring $Q$ into the form
\begin{equation}
Q=\left(\begin{array}{cc} 0&{\bf 1}\\
s  {\bf 1} &0 \end{array}\right)\,.
\end{equation}
{}{}From (\ref{eomphi}) it follows then that
\begin{equation}
\bar Q =-
W'(b)\left(\begin{array}{cc} 0&s{\bf 1}\\
{\bf 1} &0 \end{array}\right)\,.
\end{equation}
The gauge transformations that leave these matrices invariant are
given by $(U_1)^{-1} = U_2^T$ and therefore the residual gauge group
is $U(N_i)$ in this branch.

\subsubsection{Solutions with $a_i=0$}
This is the special solution of (\ref{rootsi}). The unbroken gauge
group in this branch is $U(N_i)$ and it acts on $Q$ as
\begin{equation}
Q \rightarrow U Q U^T\,.
\end{equation}
Let us consider first the case of symmetric $Q$. Since $U$ is a
general linear matrix we can choose $U$ such that $Q$ becomes the
$N_i$-dimensional unit matrix $Q={\bf 1}$.  It follows then that
$\bar{Q}=-W'(0){\bf 1}$. The gauge transformations that are left over
have to fulfill $U.U^T = {\bf 1}$ and are therefore elements of
$SO(N_i)$.

In the case where $Q$ is antisymmetric similar arguments show that $Q$
can be brought into the form
\begin{equation}
Q = \left( \begin{array}{ccc} \eps & 0&\ldots \\
0& \eps&\ldots\\
\vdots & \vdots & \ddots
\end{array}\right)\qquad \eps = \left(\begin{array}{cc} 0&1\\-1&0
\end{array}\right)\,.
\end{equation}
{} $\bar{Q}$ is determined by (\ref{eomi}) and the unbroken gauge
group is $Sp(N_i/2)$.

\subsubsection{Solutions with $W'(a_i)=0$}
In this case the equations of motion imply $Q=\bar Q =0$ and the
unbroken gauge group is $U(N_i)$ in this branch.

\subsection{The Veneziano-Yankielowicz potential}
After having established the structure of the moduli space of vacua we
derive now the leading terms in the low energy effective
superpotential. Let us take a generic vacuum with $\phi =
\mathrm{diag}(0_{N_0}, a_1 {\bf 1}_{N_1}, \dots\, , a_n {\bf 1}_{N_n},
b_1 {\bf 1}_{\tilde{N}_1}, -b_1 {\bf 1}_{\tilde{N}_1}, \dots ,b_k {\bf
  1}_{\tilde{N}_k}, -b_k {\bf 1}_{\tilde{N}_k} )$. The unbroken gauge
group is \be \prod_{i=1}^n U(N_i) \otimes \prod_{j=1}^k U(\tilde
N_j)\otimes \left\{
\begin{array}{cc} SO(N_0) & \mathrm{for}\, s=1 \\ Sp(\frac{N_0}{2})  & \mathrm{for}\, s=-1 \end{array}\right.
\ee At low energies the non-Abelian factors in the unbroken gauge
group confine and leave a $U(1)^{n+k}$ Abelian gauge group unbroken.
The effective dynamics of the gaugino condensates in the non-Abelian
factor groups is captured by the usual Veneziano-Yankielowicz
superpotential. For the unitary group factors this is \be W_{VY}^j =
S_j \log\left( \frac{\Lambda^{3N_j}_{\mathrm{low,j}}}{S_j^{N_j}}
\right)\,, \ee and for the orthogonal or symplectic factor group it is
given by
\begin{equation}
W_{VY}^s = \frac{S_0}{2} \log\left( \frac{\Lambda^{3(N_0-2s)}_{\mathrm{low,0}}}{S_0^{N_0-2s}} \right) \,.
\end{equation}
Taking all gaugino condensates into account we have
\begin{equation}
W_{eff} = \sum_{j=1}^{n+k} W^j_{VZ} + W_{VZ}^s\,,
\end{equation}
where we substitute $N_{j} = \tilde N_{j-n}$ if $j>n$.

The low energy scale can be determined by threshold matching at the
scales of the masses of the various massive $W$-bosons and matter
fields.  In particular we find for the vacuum with gauge group
$U(N_i)$
\begin{eqnarray}
\Lambda_{\mathrm{low,i}}^{3N_i} &=& \left( V''(a_i) \right)^{N_i} \prod_{j\neq
i}^n (a_i-a_j)^{-2N_j} \prod_{l=1}^k (a_i^2-b_l^2)^{-2\tilde N_l}
(a_i)^{-2N_0}\, \times\nonumber\\ & & \times\prod_{r\neq i}^n (a_i+a_r)^{N_r}
\prod_{t=1}^k (a_i^2-b_t^2)^{\tilde N_t} (a_i)^{N_0} (a_i)^{N_i+2s}
\Lambda_\mathrm{high}^{N-2s} \,.
\end{eqnarray}
The first term on the rhs of this equation comes from the fluctuations
of $\phi$ in the $N_i$'th diagonal block around the vev $a_i$. The
next terms stem from the massive off-diagonal $W$ bosons and in the
second line we collected the contributions from the matter fields $Q$
and $\bar Q$.  Analogously one can analyze the other scale matching
relations. We find:
\begin{eqnarray}\label{matchingii}
\Lambda_{\mathrm{low,i+n}}^{3\tilde N_i} &=& \left( V''(b_i)+V''(-b_i)
\right)^{\tilde N_i} \prod_{j=1}^n (b_i^2-a_j^2)^{-2 N_j} \prod_{l\neq i}^k
(b_i^2-b_l^2)^{-4\tilde N_l} (2b_i)^{-4\tilde N_i}(b_i)^{-4N_0}\times
\nonumber\\ & & \times \prod_{r=1}^n (a_r^2-b_i^2)^{N_r} \prod_{t\neq i}^k
(b_i^2-b_t^2)^{2\tilde N_t} (b_i)^{2(\tilde N_i+2s)} (b_i)^{2 N_0}
\Lambda_\mathrm{high}^{2(N-2s)} \,.
\end{eqnarray}
and
\begin{eqnarray}
\Lambda_{\mathrm{low,0}}^{\frac{3}{2}(\tilde N_0-2s)} &=& \left( V''(0)
\right)^{\frac{\tilde N_0}{2}-s} \prod_{j=1}^n (a_j)^{-2 N_j} \prod_{l=1}^k
(b_l)^{-4\tilde N_l} \prod_{r=1}^n (a_r)^{N_r}\prod_{t=1}^k (b_t)^{\tilde N_t}
\Lambda_\mathrm{high}^{N-2s} \,.
\end{eqnarray}
In (\ref{matchingii}) the factor of $2$ in the exponent of
$\Lambda_{\mathrm{high}}$ reflects the diagonal embedding of the
$U(\tilde N_i)$ gauge groups\footnote{For the threshold matching of
  the $SO/Sp$ factor group there arises the following well-known
  problem \cite{Landsteiner:1997vd}.  Projecting $SU(N)$ to $SO(N)$ or
  $Sp(\frac{N}{2})$ there are some roots which are invariant under
  both projections. However, these roots serve as long roots for
  $SO(N)$ but as short roots for $Sp(\frac{N}{2})$. This results in a
  relative factor of two in the normalization of the roots in the
  projected groups and this normalization influences the indices of
  the representations, e.g. for the adjoint representation $C_\theta =
  (\theta, \theta) g^\vee$, where $\theta$ is the highest root and
  $g^\vee$ the dual Coxeter number.}.  The higgsing by the vevs of $Q$
which breaks $U(\tilde N_j)\otimes U(\tilde N_j)$ to this diagonal
$U(\tilde N_j)$ and the $U(N_0)$ factor group to either orthogonal or
symplectic groups produces also some massive fields.  These come
however always in the multiplicities of $N=4$ multiplets and therefore
do not contribute to the threshold matching. To be specific, consider
the breaking to $SO(N_0)$. The massive fields include $W$ bosons and
components of $\phi$ that lie in the coset $U(N_0)/SO(N_0)$.  They
transform under the symmetric representation of $SO(N_0)$. In addition
$Q$ and $\bar Q$ are also symmetric in this case. Thus we count (in
${\cal N}=1$ language) $\frac{N_0}{2}(N_0 +1)$ vector multiplets and
$3\frac{N_0}{2}(N_0+1)$ chiral multiplets that receive masses at the
scale set by the vev of $Q$.

With this method we can of course obtain only an approximation to the
exact low energy superpotential.  But is it is expected to capture the
correct logarithmic behavior at $S_i=0$.  In section
\ref{sec:comp-field} we will derive this part of the superpotential
from a one-loop calculation in a matrix model.

\section{The matrix model}
In this section, we construct the matrix model which is expected to
calculate the exact superpotential of the ${\cal N}=1$ supersymmetric
$U(N)$ gauge theory with matter in the adjoint and symmetric or
antisymmetric representation.  Using direct manipulations, we shall
derive the exact loop equations of this model.  These are two
independent identities, which are respectively quadratic and cubic in
$\omega(z)$ and $\omega(-z)$, where $\omega(z)$ is the resolvent of
the model. The large $N$ limit of these relations gives algebraic
constraints on the planar limit of the averaged resolvent, which lead
to a proposal for the algebraic curve governing the dual gauge theory.
We also give evidence that the planar vacuum structure of the matrix
model agrees with the field theory.

\subsection{Construction of the model}\label{sec:constr}

Our matrix model results by performing one of two orientifold
projections on the matrix model proposed in \cite{DV2} for the $A_2$
quiver field theory.  The discussion of the covering $A_2$ quiver
theory and that of a related orbifold model can be found in Appendix
A.

We use the superpotential (\ref{treelevelsupo}) of the ${\cal N}=1$
supersymmetric gauge theory with gauge group $U(N)$, a chiral
superfield $\phi$ in the adjoint representation, a chiral superfield
$Q$ in the symmetric or antisymmetric two-tensor representation and a
chiral superfield $\bar{Q}$ in the corresponding complex conjugate
representation.  In order to define the matrix model we have to choose
a real section in matrix configuration space. In the one matrix model
this is usually done by mapping $\Phi$ to a Hermitian matrix $M$ (as
discussed in \cite{calin}, this prescription is justified if the
relevant superpotential is a polynomial of even degree). Following a
similar prescription for our models would lead to a series of
problems.  Similar to what happens in the quiver matrix models
\cite{Kostov_ade,Kostov_stat,Kostov_cft, Kharchev} and the $O(N)$
\cite{EK} we would then have to require that all eigenvalues of $M$
lie on the (strictly) positive real axis.  However, we saw in the
analysis of the classical moduli space that vacua with negative or
zero eigenvalues play an important role.  So it seems quite unnatural
from the gauge theory point of view to restrict to such matrix
configurations\footnote{A similar issue arises in quiver matrix
  models, as discussed in \cite{calin}.}. Instead, we shall consider
the matrix model: \be Z_{N,s,\eps} = {1\over \Vol U(N)} \times \int \d
M \d Q \exp\left[ -N \tr V(M) + i \tr Q^\dag M Q \right]~,
\label{eq:Z-def}
\end{equation}
where the $M$--integration is performed not over the set of Hermitian
matrices but rather over the set: \be \mathcal{M} = \big\{ M \in
\Mat_N(\mathbb{C}) \;\big|\; M-M^\dagger=2 i \eps {\bf 1} \big\}~~,
\label{setM}
\end{equation}
where $\eps$ is a small positive quantity.  We have also imposed
the condition $\bar Q = -i Q^\dagger$. Together with the shift of the
eigenvalues of $M$ into the upper half plane, this renders the
$Q$-integration finite without further restricting the range of $M$'s
eigenvalues.  Then the $Q$--integration is performed over the set: \be
\mathcal{Q} = \big\{ Q \in \Mat_N(\mathbb{C}) \;\big|\; Q^T = s Q
\;\big\}\; ,
\label{setQ}
\ee where $s=+1$ and $s=-1$ distinguishes the symmetric from the
anti-symmetric representation. The measures $\d M$, $\d Q$ are defined
through: \bea \d M &=& \prod_i \, \d M_{ii} \; \prod_{i<j} \, \d \rep
\,M_{ij} \; \d \imp \,M_{ij} \; ,
\nonumber \\
\d Q &=& \prod_{i<j} \, \d\rep \,Q_{ij} \; \d\imp \,Q_{ij} \; \Big(
\prod_i \, \d\rep \,Q_{ii} \; \d\imp \, Q_{ii} \Big)^{\delta_{s,1}}\;.
\end{eqnarray}
This model has the gauge-invariance: \bea
\label{gauge}
M&\rightarrow& UMU^\dagger\nn\\
Q&\rightarrow& UQU^T~~,
\end{eqnarray}
where $U$ is an arbitrary unitary matrix\footnote{ This can be seen as
  follows: The action and the measure $\d M$ for the adjoint field are
  obviously gauge invariant. Therefore we concentrate on the part
  coming from $\d Q$.  It is easiest to switch for a moment to an
  index notation in which the fields transform as $Q_{ij} \rightarrow
  \frac 1 2 (U_i\,^k \, U_j\,^l + s U_i\,^l \, U_j\,^k) \, Q_{kl}$,
  $\bar{Q}^{ij} \rightarrow \bar{Q}^{kl} \,\frac 1 2 (U^\dagger_k\,^i
  \, U^\dagger_l\,^j + s U^\dagger_l\,^i \, U^\dagger_k\,^j ) $ We
  took care explicitly of the symmetry properties of $Q$ and $\bar Q
  $.  The measure $\d Q$ picks up the product of the determinants of
  the transformation matrices defined in the previous equations. Since
  the product of the determinants is the determinant of the product we
  compute the product of the transformation matrices: $ \frac 1 2
  (U_i\,^k \, U_j\,^l + s U_i\,^l \, U_j\,^k) .  \frac 1 2
  (U^\dagger_k\,^m \, U^\dagger_l\,^n + s U^\dagger_l\,^m \,
  U^\dagger_k\,^n ) = \frac 1 2 ( \delta^m_i\, \delta^n_j +
  s\delta^n_i\, \delta^m_j) $ This is just the identity in the
  symmetric and antisymmetric representation respectively and of
  course the corresponding determinant is one.}.  Note that we have
included the inverse volume of the gauge group into the definition of
the partition function \erf{eq:Z-def}.  Also note that we are rather
valiant about the convergence of (\ref{eq:Z-def}), which is assured
---with our choice of integration manifold --- only provided that $W$
is a polynomial of {\em even} degree. As in \cite{calin}, a consistent
construction for odd degree potentials would require that we constrain
the eigenvalues of $M$ to lie on a certain path in the complex plane,
whose asymptotic behavior is determined by the leading coefficient of
$W$. In the present paper, we shall ignore this and related issues, a
complete treatment of which requires the full machinery of
\cite{calin}.

Performing the Gaussian integral over $Q$ yields \be
\int_{\mathcal{Q}}\!\!\!dQ \; e^{ i \; \tr Q^\dagger M Q} = \big(i
\pi\big)^{N(N+s)/2} \; \prod_{i<j} \frac{1}{\la_i+\la_j} \;
\Big(\prod_i \frac{1}{\la_i} \Big)^{\delta_{s,1}}
\end{equation}
where $\la_1,\dots,\la_N$ are the eigenvalues of the matrix $M$.

Next we rewrite the integral over $M$ in \erf{eq:Z-def} in terms of
eigenvalues of $M$: \be Z_{N,\eps,\delta} = C_{N,s} \; \tilde
Z_{N,s,\eps} \;,
\end{equation}
with \be \tilde Z_{N,s,\eps} = \int_{\R +i\eps\ } \prod_{k=1}^N d\la_k
\, e^{-N^2 S} \quad {\rm and} \quad C_{N,s} = {2^{-N^2/2+N} \over \Vol
  (U(1)^N\times S_N)} \big(i\pi\big)^{N(N+s)/2} ~~,
\label{eq:Zt}
\end{equation}
where the `effective' action $S(\la_1,\dots,\la_N)$ in the eigenvalue
representation is: \be S = \frac{1}{N} \sum_k V(\la_k) +
\frac{s}{2N^2} \sum_k \ln\la_k - \frac{1}{2N^2} \sum_{k \neq l}
\ln(\la_k{-}\la_l)^2 + \frac{1}{2N^2} \sum_{k,l} \ln(\la_k{+}\la_l)
\;.
\label{eq:action}
\end{equation}

In the limit $\eps\rightarrow 0$ (which we shall ultimately take
below), the factors
$e^{-\ln(\lambda_k+\lambda_l)}=\frac{1}{\lambda_k+\lambda_l}$ in
(\ref{eq:Zt}) produce well defined distributions according to the
Sokhotsky formula: \be
\label{Sokh}
\frac{1}{\lambda_k+\lambda_l + i 0^+} =
\mathcal{P}\left(\frac{1}{\lambda_k+\lambda_l}\right) - i \pi
\delta(\lambda_k+\lambda_l)\;, \ee where we took the limiting
eigenvalues to be real in order to make the $i\eps$ prescription
explicit and where $\mathcal{P}$ denotes the principal value.
Substituting (\ref{Sokh}) into (\ref{eq:Zt}) (before exponentiating
the Vandermonde factors) leads to an expression for the
$\eps\rightarrow 0^+$ limit of ${\tilde Z}$ as a sum over
`reduced' integrals. This gives a limiting statistical ensemble, which
can viewed as the orientifold of the limiting ensemble extracted in
\cite{calin} for the holomorphic $A_2$ model.

The potentials $V(x)$ we want to consider are of the form \be V(x)
\;=\; t_{-1} \ln(x) + W(x) \qquad {\rm where} \quad W(x) \;=\;
\sum_{k=0}^d \frac{t_k}{k{+}1} \, x^{k+1} \;.
  \label{eq:potential}
\end{equation}
{}From \erf{eq:action} we see that the effect of integrating only over
symmetric or anti-symmetric $Q$ is a logarithmic correction to the
potential of order $1/N$. If we introduce \be U(x) = V(x) +
\frac{s}{2N} \ln(x)
\label{eq:U-def}
\end{equation}
the action $S$ depends on $s$ only implicitly through $U(x)$.  For the
analysis of the matrix model it is useful to further introduce the
following quantities.  The expectation value of a function
$\mathcal{O}(\la_1,\dots,\la_N)$ is given by \be \langle \mathcal{O}
\rangle = \frac{1}{\tilde Z} \int_{\R +i\eps\ } \prod_{k=1}^N d\la_k
\; \mathcal{O}(\la_1,\dots,\la_N) \; e^{-N^2 S}
\end{equation}
and the eigenvalue density $\rho(\la)$ and the resolvent $\omega(z)$
are defined to be \be \rho(\la) = \frac1N \sum_k \delta(\la{-}\la_k)
\quad{\rm and}\quad \om(z) = \frac1N \sum_k \frac{1}{z-\la_k} \;.
\end{equation}
The two quantities are related by \be \rho(\la) = \lim_{\nu\rightarrow
  0} \frac{ \om(\la{+}i\nu) - \om(\la{-}i\nu) }{2 \pi i} \quad ,
\qquad \om(z) = \int_{\R +i\eps\ } \!\!\! d\la \,
\frac{\rho(\la)}{z{-}\la} ~.
\end{equation}

\subsection{The microcanonical ensemble}

The framework of Dijkgraaf-Vafa requires that the model obeys certain
filling fraction constraints.  In \cite{DV1}, such conditions were
imposed only on the large $N$ microcanonical generating function,
which is insufficient in our case since we will have to consider
$O(1/N)$ corrections as can be seen from (\ref{eq:U-def}).
This requires that we impose such constraints on the finite $N$
microcanonical generating function, rather than on its large $N$
counterpart. The relevant constraints are easiest to formulate by
employing a microcanonical ensemble. As we shall see below, the
original path integral defines a (grand) canonical ensemble at zero
chemical potentials. This allows one to recover the microcanonical
generating function by introducing non-vanishing chemical potentials
(which are canonically conjugate to the filling fractions) and then
performing a Legendre transform to replace the former by the latter.

\subsubsection{The (grand) 
  canonical partition function associated with a collection of
  intervals}

Let us cover the displaced real line $\R + i \eps\ $ with disjoint
nonempty line segments $\Delta_\alpha$ subject to the condition: \be
\Delta_1 \cup \cdots \cup \Delta_r = \R+i \eps \;.
\end{equation}
We shall let $\chi_\alpha$ denote the characteristic function of
$\Delta_\alpha$, and consider the filling fractions $f_\alpha$ of
$\Delta_\alpha$: \be f_\alpha = \frac1N \sum_k \chi_\alpha(\la_k)~~.
\end{equation}
This gives the expression of $f_\alpha$ in terms of $\om(z)$: \be
f_\alpha =\int{d\lambda \rho(\lambda)\chi_\alpha(\lambda)}=
\int_{\Delta_\alpha}{d\lambda \rho(\lambda)}=
\oint_{\gamma_\alpha}{\frac{dz}{2\pi i}\omega(z)}~~,
\end{equation}
Picking chemical potentials $\mu_\alpha$, we consider the (grand)
canonical ensemble associated to our collection of
intervals\footnote{In the following $\mu,S$ may stand for $r$-tuples
  and $t$ for $(t_{-1},t_0,\dots,t_d)$.}: \be {\cal Z}(t, \mu) =
\int_{\R + i \eps\ } \prod_{k=1}^N d\la_k \, e^{-N^2 S_\mu } \quad
{\rm where} \quad S_\mu = S + \sum_{\alpha=1}^r \mu_\alpha f_\alpha ~.
\end{equation}
The original partition function results by setting $\mu_\alpha=0$.
Introducing the (grand) canonical generating function: \be {\cal
  F}(t,\mu) \;=\; -\frac{1}{N^2}\ln {\cal Z}(t,\mu)~~,
\end{equation}
we have the standard relation: \be
\label{cfractions}
\frac{\partial}{\partial \mu_\alpha}{\cal F}=\langle f_\alpha\rangle =
\oint_{\gamma_\alpha}{\frac{dz}{2\pi i}\langle \omega(z)\rangle}~~.
\end{equation}
In this subsection, the brackets $\langle \dots \rangle$ always denote
the expectation value taken in the (grand) canonical ensemble.  Since
the union of $\Delta_\alpha$ covers the whole integration range, the
expectation values of the filling fractions fulfill the constraint:
\be 
\sum_{\alpha=1}^r \langle f_\alpha \rangle = 1 \;.
  \label{norm}
\end{equation}

\subsubsection{The microcanonical generating function}

Following standard statistical mechanics procedure, we define: \be
\label{S_def}
S_\alpha:=\frac{\partial}{\partial\mu_\alpha}{\cal F}
\end{equation}
and perform a Legendre transform to extract the microcanonical
generating function: \be
\label{F_def}
F(t, S):=\sum_{\alpha = 1}^r S_\alpha \mu_\alpha( t, S) -{\cal
  F}(t,\mu(t, S))~~.
\end{equation}
In this relation, $\mu_\alpha$ are expressed in terms of $ t$ and $S$
by solving equations (\ref{S_def}). The constraint (\ref{norm}) shows
that $S_\alpha$ are related through: \be
\label{Snorm}
\sum_{\alpha=1}^r{S_\alpha}=1~~,
\end{equation}
so we can take $S_1\dots S_{r-1}$ to be the independent variables.
Then equations (\ref{S_def}) express $\mu_\alpha$ as functions of $t$
and these coordinates, and equation (\ref{F_def}) implies: \be
\label{quantum_periods}
\mu_\alpha-\mu_r=\frac{\partial F}{\partial
  S_\alpha}~{\rm~for~}\alpha= 1,\dots,r-1~~.
\end{equation}
Note that $\mu_\alpha$ are only determined up to a common constant
shift; this is due to the constraint (\ref{Snorm}) on $S_\alpha$.

Working with $F( t, S)$ amounts to fixing the expectation values of
the filling fractions by imposing the {\em quantum} constraint
(\ref{S_def}): \be
\label{fconstraint}
\langle f_\alpha \rangle= \oint_{\gamma_\alpha}{\frac{dz}{2\pi
    i}\langle \omega(z)\rangle}=S_\alpha~~,
\end{equation}
with $S_\alpha$ treated as fixed parameters.  This gives a meaning to
the procedure of \cite{DV1} beyond the large $N$ limit.

\subsection{The quadratic and cubic loop equations}
\label{subs:le}
The classical equations of motion following from the action
\erf{eq:action} are $\partial S/\partial\la_k = 0$, where: \be N
\frac{\partial S}{\partial\la_k} = U'(\la_k) - \frac1N \sum_{l(\neq
  k)} \frac{2}{\la_k{-}\la_l} + \frac1N \sum_l \frac{1}{\la_k{+}\la_l}
\;.
  \label{eq:EOM}
\end{equation}
The action \erf{eq:action} describes a system of charged particles
moving along $\mathbb{R} + i \eps$, together with a set of mirror
charges of opposite sign.  For $s{=}{-}1$, each particle interacts
with all mirror charges excluding its own, while for $s{=}1$ it is
also attracted to its own mirror image.

The partial derivatives \erf{eq:EOM} will be used below to obtain a
set of Ward identities for the matrix model. In turn, we shall use
these identities in order to derive two loop equations for resolvent
$\om(z)$.  Consider the integral: \be \frac{1}{N^2} \, \frac{1}{\tilde
  Z} \int_{\mathbb{R}+i\eps} \prod_{i=1}^N d\la_i \; \sum_k
\frac{\partial}{\partial \la_k} \Big( \psi_k(\la_1,\dots,\la_N) \;
e^{-N^2 S} \Big) \;= 0 \;
\label{eq:frel-1}
\end{equation}
Differentiating shows that \erf{eq:frel-1} is equivalent to the Ward
identity \be \frac1N \sum_k \Big\langle \Big( U'(\la_k) - \frac
1N\sum_{l(\neq k)} \frac{2}{\la_k{-}\la_l} + \frac1N \sum_l
\frac{1}{\la_k{+}\la_l} \Big) \psi_k - \frac1N \frac{\partial
  \psi_k}{\partial \la_k} \Big\rangle \;=\; 0\;.
\label{eq:Ward}
\ee

\subsubsection{The quadratic loop equation}

The quadratic loop equation is obtained by setting $\psi_k =
(z{-}\la_k)^{-1}$. Together with the identity \be 2 \sum_{k\neq l}
\frac{1}{\la_k{-}\la_l}\, \frac{1}{z{-}\la_k} + \sum_k
\frac{1}{(z{-}\la_k)^2} = \sum_{k,l}
\frac{1}{z{-}\la_k}\,\frac{1}{z{-}\la_l}
 \label{eq:2lam-ident}
\end{equation}
one obtains the relation \be \Big\langle \om(z)^2 - \frac1N\sum_k
\frac{U'(\la_k)}{z{-}\la_k} - \frac{1}{N^2} \sum_{k,l}
\frac{1}{\la_k{+}\la_l} \frac{1}{z{-}\la_k} \Big\rangle \;=\; 0\;.
\label{eq:half-q-loop}
\end{equation}
Adding to \erf{eq:half-q-loop} the same equation with $z$ replaced by
$-z$ and absorbing the dependence on the potential into $f(\pm z)$
leads to the {\em quadratic loop equation}: \be \big\langle \om(z)^2 +
\om(z)\om(-z) + \om(-z)^2 - U'(z) \om(z) - U'(-z) \om(-z) + f(z) +
f(-z) \big\rangle = 0 \;.
\label{eq:l2}
\ee In deriving this we used the identity: \be \sum_{k,l}
\frac{1}{\la_k{+}\la_l} \, \frac{1}{z{-}\la_k} + \sum_{k,l}
\frac{1}{\la_k{+}\la_l} \, \frac{1}{-z{-}\la_k} = - \sum_{k,l}
\frac{1}{z{-}\la_k} \, \frac{1}{-z{-}\la_l}
\label{eq:other_identity}
\end{equation}
and introduced the quantity: \be f(z) = \frac 1N \sum_k \frac{U'(z) -
  U'(\la_k)}{z-\la_k} \;.
  \label{eq:f-fbar}
\end{equation}
In the analysis of the large $N$ Riemann surface below we will need to
know the properties of $f(z)$ in more detail. {}From the definition of
$U,V,W$ in \erf{eq:potential} and \erf{eq:U-def} it follows that
$f(z)$ can be written: \be f(z) = \bar f(z) - \phi \cdot
\frac{t_{-1}+s/(2N)}z \qquad {\rm where} \quad \phi = \frac1N \sum_k
\la_k^{-1} = -\om(0) \;.
\label{eq:f-decomp}
\end{equation}
The function $\bar f(z)$ is a polynomial in $z$ defined through: \be
\bar f(z) = \frac1N \sum_k \frac{W'(z) - W'(\la_k)}{z-\la_k}\;.
  \label{eq:fbar-deg}
\end{equation}
Since $\om(z) \sim 1/z$ as $z\rightarrow \infty$, one finds that the
large $z$ behavior of $\bar f$ is $\bar f(z) \sim t_d z^{d-1}$. In
particular the polynomial $\bar f(z)$ has degree $d{-}1$.

\subsubsection{The cubic loop equation}

The cubic relation for $\om(z)$ results from substituting \bea \psi_k
&=& \frac1N\sum_m \frac{1}{(\la_m{+}\la_k)(z{-}\la_k)}
\qquad {\rm ,~so~that:} \nn\\
\frac{\partial \psi_k}{\partial \la_k} &=& \frac1N\sum_m
\frac{1}{(\la_m{+}\la_k)(z{-}\la_k)} \Big(\frac{1}{z{-}\la_k} -
\frac{1}{\la_m{+}\la_k}\Big) + \frac1N \, \frac1{4 \,\la_k^{\,2} \,
  (z{-}\la_k)}
  \label{eq:cub-1}
\end{eqnarray}
into the Ward identity \erf{eq:Ward}.  To proceed we make use of the
relations: \bea &&-\frac1{N^3}\sum_{m,k,l(\neq k)}
\frac{2}{(\la_k{-}\la_l)(\la_m{+}\la_k)(z{-}\la_k)}
\nonumber\\
&& \qquad = \; \om(z)^2\om(-z) + \frac1{N^3}\sum_{k,m}
\frac{1}{(\la_m{+}\la_k)(z{-}\la_k)} \Big(\frac{1}{z{-}\la_k} -
\frac{1}{\la_m{+}\la_k}\Big)
\nonumber\\
&& \qquad\qquad + \frac1{N^3}\sum_{m,k,l}
\frac{1}{(\la_m{+}\la_k)(\la_l{+}\la_k)(z{+}\la_k)}
\label{eq:cub-2}
\end{eqnarray}
and \be \frac1N \sum_k \frac{1}{\la_k^{\,2}\,(z{-}\la_k)} \; = \;
\frac{\om(z){-}\om(0)}{z^2} \;+\; \frac1z\,\frac1N\sum_k
\frac1{\la_k^{\,2}}
\label{eq:cub-2b}
\end{equation}
Equations \erf{eq:cub-2} and \erf{eq:cub-2b} follow from the partial
fraction decompositions: \bea && \frac{1}{(z{-}a)(z{-}b)(z{-}c)}=
\frac{1}{(b{-}a)(c{-}a)(z{-}a)}+ \frac{1}{(c{-}b)(a{-}b)(z{-}b)}+
\frac{1}{(a{-}c)(b{-}c)(z{-}c)}
\nonumber \\
&& \frac{1}{(z{-}a)^2(z{-}b)}= \frac{1}{(a{-}b)^2(z{-}b)}-
\frac{1}{(a{-}b)^2(z{-}a)}+ \frac{1}{(a{-}b)(z{-}a)^2}
\label{cubfrac}
\end{eqnarray}
When substituting \erf{eq:cub-1}, \erf{eq:cub-2}, \erf{eq:cub-2b} into
\erf{eq:Ward} some of the terms cancel and we are left with \bea
&&\Big\langle \om(z)^2\om(-z) + \frac{1}{N^2} \sum_{k,m}
\frac{U'(\la_k)}{(\la_m{+}\la_k)(z{-}\la_k)} + \frac{1}{N^3}
\sum_{m,k,l} \frac{1}{(\la_l{+}\la_k)(\la_m{+}\la_k)}
\Big(\frac{1}{z{-}\la_k} - \frac{1}{-z{-}\la_k}\Big)
\nonumber\\
&&\qquad - \frac1{N^2}
\Big(\frac{\om(z){-}\om(0)}{4\,z^2}+\frac{1}{4z} \frac1N\sum_k
\la_k^{-2} \Big) \Big\rangle \;=\; 0 \quad
\label{eq:half-c-loop}
\end{eqnarray}
As in the quadratic case we can add to this the same equation with $z$
replaced by $-z$. This removes the triple sum and gives \bea
&&\Big\langle \om(z)^2\om(-z) + \om(z)\om(-z)^2 + \frac{1}{N^2}
\sum_{k,m} \frac{U'(\la_k)}{\la_m{+}\la_k} \Big( \frac{1}{z{-}\la_k} -
\frac{1}{z{+}\la_k} \Big)
\nonumber \\
&& \qquad - \frac1{N^2} \,\frac{\om(z){+}\om(-z){-}2\om(0)}{4\,z^2}
\Big\rangle \;=\; 0
  \label{eq:aux-1st-order}
\end{eqnarray}
To proceed we introduce the quantity \be g(z) =
\frac{1}{N^2}\sum_{k,l}
\frac{U'(z)-U'(\la_k)}{(\la_l{+}\la_k)(z{-}\la_k)} \;.
\end{equation}
As with $f(z)$ in \erf{eq:f-decomp}, for later use we note that $g(z)$
can decomposed into a polynomial part $\bar g(z)$ and a pole $1/z$ as
\bea && g(z) = \bar g(z) - \gamma \cdot \frac{t_{-1}+s/(2N)}z
\qquad {\rm where} \nn \\
&& \bar g(z) = \frac{1}{N^2}\sum_{k,l}
\frac{W'(z)-W'(\la_k)}{(\la_l{+}\la_k)(z{-}\la_k)} \quad {\rm and}
\quad \gamma = \frac{1}{N^2}\sum_{k,l} \frac1{\la_k(\la_k{+}\la_l)} =
\frac{\om(0)^2}2 \qquad
  \label{eq:gbar-deg}
\end{eqnarray}
As before, since $W'(z)$ has degree $d$ the large $z$ behavior of the
polynomial $\bar g(z)$ is $({\rm const}) \cdot z^{d-1}$, so that $\bar
g(z)$ has degree $d{-}1$. The expression $\gamma = \om(0)^2/2$ has
been obtained by evaluating the identity \erf{eq:other_identity} at
$z{=}0$.

Using \erf{eq:half-q-loop}, one finds the following relation for the
expectation value of $g(z)$: \be \big\langle g(z) \big\rangle =
\Big\langle - \frac{1}{N^2}\sum_{k,l}
\frac{U'(\la_k)}{(\la_l{+}\la_k)(z{-}\la_k)} + U'(z) \Big( \om(z)^2 +
f(z) - U'(z) \om(z) \Big) \Big\rangle
\label{eq:cub-3}
\end{equation}
Substituting \erf{eq:cub-3} into \erf{eq:aux-1st-order} we obtain the
{\em cubic loop equation}: \bea &&\big\langle \; \om(z)^2\om(-z) -
g(z) + U'(z) \big( \om(z)^2 + f(z) - U'(z) \om(z) \big)
\;+\; (z\leftrightarrow-z) \; \big\rangle \nonumber\\[3pt]
&& ~~ - \; \frac1{N^2} \,\frac{\langle
  \om(z){+}\om(-z){-}2\om(0)\rangle}{4\,z^2} \;=\; 0~~.  \qquad
\label{eq:l3}
\end{eqnarray}

\subsubsection{Loop equations in terms of contour integrals}

The two loop equations \erf{eq:l2} and \erf{eq:l3} can be presented in
a more compact form when using contour integrals.  This form can be
obtained as follows. Let $\gamma$ be a contour that encircles all the
eigenvalues but not the point $z$ and not the poles of $\om(-z)$.  One
can verify the two identities \bea &&\oint_{\gamma} \frac{dx}{2\pi i}
\frac{2x U'(x)}{z^2{-}x^2} \big\langle \om(x)\big\rangle \;=\;
\frac{1}{N} \sum_k \Big\langle \frac{2\lambda_k
  U'(\lambda_k)}{z^2{-}\lambda_k^2}
  \Big\rangle \quad, 
  \nonumber\\
  &&\oint_{\gamma} \frac{dx}{2\pi i} \frac{2x U'(x)}{z^2{-}x^2}
  \big\langle \om(x)\om(-x)\big\rangle \;=\; \frac{1}{N^2} \sum_{k,l}
  \Big\langle \frac{2\lambda_k U'(\lambda_k)}{z^2{-}\lambda_k^2}
  \frac{-1}{\la_k{+}\la_l}\;.  \Big\rangle
\end{eqnarray}
and insert them into the equations \erf{eq:l2} and
\erf{eq:aux-1st-order}.  This results in the following constraints,
which are equivalent with the loop equations \erf{eq:l2} and
\erf{eq:l3}: \bea
  \label{eq:intthirdloop}
  &&\big\langle \om(z)^2+\omega(z)\omega(-z)+\om(-z)^2 \big\rangle
  \;=\; \oint_{\gamma} \frac{dx}{2\pi i} \, \frac{2x U'(x)}{z^2{-}x^2}
  \, \big\langle \om(x)\big\rangle
  \quad , \qquad \nn\\[8pt]
  &&\big\langle \om(z)^2\om(-z) + \om(z)\om(-z)^2 \big\rangle - \;
  \frac1{N^2}
  \,\frac{\langle\om(z){+}\om(-z){-}2\om(0)\rangle}{4\,z^2}
  \\[4pt]
  && \qquad \;=\; \oint_{\gamma} \frac{dx}{2\pi i} \frac{2x
    U'(x)}{z^2{-}x^2} \big\langle \om(x)\om(-x)\big\rangle \nn~~.
  \eea

\subsection{The large $N$ Riemann surface}\label{sec:largeN}

In this section we shall take the large $N$ limit of the loop
equations for the orientifold with symmetric and antisymmetric matter,
thus obtaining two polynomial constraints on the planar limit
$\omega_0$ of the averaged resolvent \footnote{These constraints also
  follow from the loop equations of the $A_2$-quiver, which are given
  in Appendix \ref{sec:large-N-A2}.}, which is the leading term in the
large $N$ expansion: \be \langle \om(z) \rangle = \om_{0}(z) +{1\over
  N} \om_{1}+ O(1/N^2) \ee These large N constraints on
$\om_{0}$ define a Riemann surface, similar to what happens in the
much better studied case of one-matrix models.  The main difference is
that the Riemann surface governing our models is not hyperelliptic.
Rather, it will turn out to be a triple cover of the $z$-plane, one of
whose branches gives $\om_{0}(z)$.

\subsubsection{The loop equations at large $N$}

The large $N$ limit of the loop equations \erf{eq:l2} and \erf{eq:l3}
gives: \bea && \om_{0}(z)^2 + \om_{0}(z)\om_{0}(-z) + \om_{0}(-z)^2 -
V'(z) \om_{0}(z) - V'(-z) \om_{0}(-z) + f(z) + f(-z) \;=\; 0
\nonumber\\
&& \om_{0}(z)^2 \om_{0}(-z) - g(z) + V'(z) \big( \om_{0}(z)^2 + f(z) -
V'(z) \om_{0}(z) \big) \;+\; (z\leftrightarrow-z) \;=\; 0
\label{eq:genus0}
\eea It is convenient to introduce the shifted variable $u(z)$ through
the relations: \be
\label{shiftomega}
\omega_0(z)=u(z)-t(z)~~,~~\omega_0(-z)=u(-z)-t(-z)~~, \end{equation}
where \be
\label{tdef}
t(z)=\frac{-2V'(z)+V'(-z)}{3}~~, ~~t(-z)=\frac{-2V'(-z)+V'(z)}{3}~~.
\end{equation} Under this translation, equations \erf{eq:genus0}
become: \bea
\label{Viete}
u(z)^2+u(z)u(-z)+u(-z)^2&=&p(z)\nn\\ 
u(z)^2u(-z)+u(z)u(-z)^2~~~~&=&-q(z)~~,
\end{eqnarray} where:
\bea
p(z)&=&t(z)^2+t(z)t(-z)+t(-z)^2-f(z)-f(-z)\nn\\
q(z)&=&-t(z)t(-z)\left[t(z)+t(-z)\right]+t(z)f(-z)+t(-z)f(z)-g(z)-g(-z)~~.
\label{pq}
\end{eqnarray}
Equations (\ref{Viete}) are the Viete relations for the roots
$u_0(z):=u(z), u_1(z):=u(-z)$ and $u_2(z):=-u_0(z)-u_1(z)$ of the
cubic: \be
\label{curve}
\prod_{i=0}^2 (u(z)-u_i(z)) \;=\; u^3-p(z)u-q(z)\;=\;0~~.
\end{equation} 
Therefore, the quantities $u_0(z), u_1(z)$ and $u_2(z)$ are the three
branches of the affine curve (\ref{curve}), when the latter is viewed
as a triple cover of the complex $z$-plane.

Let us index the branches such that: \be
\label{enumeration}
\omega_0(z)=u_0(z)-t(z) \quad{\rm and}\quad \omega_0(-z)=u_1(z)-t(-z)
\;.
\end{equation}
It is clear from (\ref{pq}) that the functions $p(z)$ and $q(z)$ are
even. Hence the curve (\ref{curve}) admits the automorphism: \be
\nu(z,u)=(-z,u) \ ,
  \label{eq:u-aut}
\end{equation}
which permutes the sheets $u_0$ and $u_1$, while stabilizing the third
sheet.  Using (\ref{pq}), our curve can be written: \bea
&&\big(u-t(z)\big)\big(u-t(-z)\big)\big(u+t(z)+t(-z)\big) \nn\\
&&\qquad +\; \big[\,f(z)+f(-z)\,\big]u \;+\;
\big[\,2g_{ev}(z)-t(z)f(-z)-t(-z)f(z)\,\big] \;=\; 0~~,
\label{curve_def}
\end{eqnarray} 
where: \be g_{ev}(z):=\frac{1}{2}\left[g(z)+g(-z)\right] =
\frac{1}{2}\left[\bar g(z)+\bar g(-z)\right]
\end{equation} is the even part
of $g$. For $f=g\equiv 0$, the defining equation reduces to: \be
(u-t_0(z))(u-t_1(z))(u-t_2(z))=0~~,
\label{curve0}
\end{equation} 
where $t_0(z):=t(z),t_1(z):=t(-z),t_2(z)=-t_0(z)-t_1(z)$.
Correspondingly, the branches are given by $u_j=t_j$ in this limit.
The general curve (\ref{curve_def}) is a deformation of
(\ref{curve0}), parameterized by the coefficients of $f$ and $g_{ev}$.

The number of independent coefficients in $f$ and $g_{ev}$ is
constrained by the matrix model.  Recall from the discussion of
equations \erf{eq:fbar-deg} and \erf{eq:gbar-deg} that the polynomials
$\bar f(z)$ and $\bar g(z)$ have degree $d{-}1$ if the polynomial part
$W(x)$ of the potential has degree $d{+}1$.  Thus $\bar f(z)$ depends
on $d$ complex parameters while the even polynomial $g_{ev}$ depends
on $\delta{+}1$ parameters, where \be \delta =
\big[\tfrac{d{-}1}2\big] \;.
\end{equation}
In the large $N$ limit the function $f(z)$ in \erf{eq:f-fbar} has the
form $f(z) = \bar f(z) - \phi t_{-1}/z$, so it depends on the
coefficients of $\bar f(z)$ as well as on $\phi$, if the logarithmic
term is present in the potential.  Altogether, we have: \be \#(\,{\rm
  coefficients~in~}f,\,g_{ev}\,) = \begin{cases}
  d + \delta + 1 & ; t_{-1} = 0 \qquad \qquad \qquad a.) \\
  d + \delta + 2 & ; t_{-1} \neq 0 \qquad \qquad \qquad b.)
  \end{cases}
  \label{eq:f-gev-coeff}
\end{equation}

\subsection{The Riemann surface in the absence of a logarithmic deformation}

\subsubsection{General description and parameter count}

In the absence of a logarithmic deformation ($t_{-1}=0$), the large
$N$ algebraic curve is given by (\ref{curve}) with $V(z)=W(z)$.  We
shall start by counting its parameters and periods.  For this, we must
describe the branching structure of this curve and of its
deformations.  For a generic deformation $(f,g)$, we shall find
$d+\delta+1$ independent periods, in agreement with the parameter
count for $g_{ev}$ and $f$ performed in the previous subsection.

Let us start with the classical curve (\ref{curve0}) and analyze its
branching. Writing $W'(z)=\sum_{m=0}^d{t_mz^m}$, we have: \be
W'_{odd}(z):=\frac{1}{2}[W'(z)-W'(-z)]=\sum_{k=0}^{\delta}
{t_{2k+1}z^{2k+1}}=zv(z^2)~~, \ee where we defined
$v(x):=\sum_{k=0}^{\delta}{t_{2k+1}x^k}$.  Consider the
factorizations: \be W'(z)=t_d(z-z_1^+)\dots (z-z_d^+)~~\rightarrow
W'(-z)=t_d (-1)^d(z-z_1^-)\dots (z-z_d^-) \ee with $z_i^-:=-z_i^+$ and
\be v(x)=t_{2\delta+1}(x-x_1)\dots (x-x_\delta)~~.  \ee

The second relation implies the factorization: \be
W'_{odd}(z)=t_{2\delta+1} z(z^2-x_1)\dots (z^2-x_\delta)=
t_{2\delta+1} z (z-{\tilde z}_1^+)(z-{\tilde z}_1^-)\dots (z-{\tilde
  z}_\delta^+)(z-{\tilde z}_\delta^-)~~, \ee where ${\tilde
  z}_j^{\pm}=\pm \sqrt{x_j}$.

Then the curve (\ref{curve0}) has ordinary double points at those
values of $z$ where $t_i(z)=t_j(z)$ for some $0\leq i <j \leq 2$.  Let
us assume that all roots $z^+_i, z^-_i,{\tilde z}^+_j$ and ${\tilde
  z}_j^-$ are mutually distinct and nonzero (this is the generic
case).  Then: \bea t_0(z)&=&t_2(z)\Leftrightarrow W'(z)=0
\Leftrightarrow z\in\{z_1^+\dots z_d^+\}~~\nn\\
t_1(z)&=&t_2(z)\Leftrightarrow W'(-z)=0
\Leftrightarrow z\in\{z_1^-\dots z_d^-\}~~\\
t_0(z)&=&t_1(z)\Leftrightarrow W'_{odd}(z)=0 \Leftrightarrow z\in
\{{\tilde z}_1^-\dots {\tilde z}_\delta^-, 0, {\tilde z}_1^+\dots
{\tilde z}_\delta^+\}~~.\nn \eea

When turning on generic deformations $f,g$ to reach the curve
(\ref{curve}), all of these double points will split into cuts: \bea
z_i^+\rightarrow [a_i,b_i]:=I_i~, &~z_i^-\rightarrow
[-b_i,-a_i]:=I_{-i}~,
&~i=1,\ldots, d\nn\\
{\tilde z}_j^+\rightarrow [{\tilde a}_j,{\tilde b}_j]:={\tilde I}_j~,
&~{\tilde z}_j^-\rightarrow [-{\tilde b}_j,-{\tilde a}_j]:={\tilde
  I}_{-j}~,
&~j=1,\ldots ,\delta \nonumber\\
\tilde z_0 \rightarrow [\tilde a_0, \tilde b_0] := \tilde I_0~, &~{\rm
  with} ~ \tilde a_0 = - \tilde b_0~. &~
\label{genericpert}\eea
This splitting is symmetric since the allowed deformations must
preserve the symmetry of our curve. We obtain $2d$ cuts of type $I_i$
and $2\delta+1$ cuts of type ${\tilde I}_j$.

The deformed curve (\ref{curve}) has a multiple point at $z=\infty$.
It follows that an appropriate deformation of its normalization will
have two supplementary cuts. Applying the Hurwitz formula for this
normalized and deformed curve, we obtain its genus $g$: \be
\label{RH}
g=\frac{1}{2}(4\delta +4d+6)-3+1=2(d+\delta)+1~~.  \ee This is one
unit greater than the arithmetic genus of the curve (\ref{curve}).

\begin{figure}[bt]
\begin{center}
  \scalebox{0.7}{\input{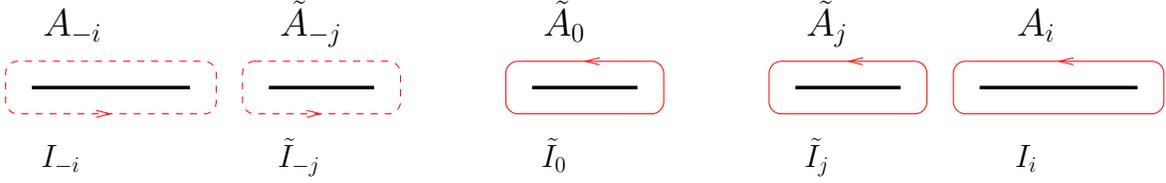}}
\end{center}
\caption{The cycles $A_i$ and ${\tilde A}_j$.
  Solid lines lie on the sheet $u_0(z)$, while dashed lines lie on
  $u_1(z)$. Under the automorphism $\nu$ the cycles transform as
  $\nu(A_i) = A_{-i}$, $\nu(\tilde A_j) = \tilde A_{-j}$ and
  $\nu(\tilde A_0) = - \tilde A_0$.}
\label{Acycles}
\end{figure}

Choose cycles $A_i$, $\tilde A_j$ around the cuts $I_i$, $\tilde I_j$
as indicated in figure \ref{Acycles}.  Due to the $\Z_2$ invariant
perturbation of the cuts and the induced action $\nu^*(\la)=-\la$ of
(\ref{eq:u-aut}) on the meromorphic differential \be \lambda={1\over 2
  \pi i} \, u \, dz , \ee the periods integrals of this `regularized'
curve along $A_{-i},\tilde A_{-j}$ are minus the ones along
$A_{i},\tilde A_{j}$ for $i=1\dots d$ and $j=1\dots\delta$,
while $\int_{\tilde A_0} \la$ is invariant. Therefore, we can choose
$\int_{A_i}{\lambda}$ and $\int_{{\tilde A}_j}{\lambda}$ (with
$i=1\dots d$ and $j=0\dots \delta$) to be independent periods. Hence their
number agrees with the parameter count for $g_{ev}$ and
$f$.  In (\ref{genericpert}) we assumed a generic situation, which can be
justified from the double cover. In that case, the existence of $3d$
independent $3$-cycles in the Calabi-Yau geometry has been established
\cite{Cachazo:2001gh}. By the classical results of Tian and Todorov
\cite{tt,tt2}, this is equivalent to $3d$ unobstructed complex
structure parameters, which descend to the Riemann surface. The
argument above can be viewed as establishing consistency of the $\Z_2$
projection on the parameters and periods of the Riemann surface.

\subsubsection{Physical meaning of the cuts}\label{sec:phys}

We next discuss the interpretation of the cuts in terms of the planar
eigenvalue distribution of the matrix model. Remember that
$\omega_0(z)=\int{d\lambda \frac{\rho(\lambda)}{z-\lambda}}$. {}From
\erf{enumeration} we see that the cuts of $\omega_0(z)$ coincide with
those of $u_0(z)$. The fact that $u_0$ has branch cuts along $I_i$ and
${\tilde I}_j$ (with $i=1\dots d$ and $j=-\delta\dots \delta$)
requires that $\rho(\lambda)$ be non-vanishing along {\em each} of
these cuts.  This implies that $\omega_0(-z)$ will have cuts on the
reflected loci $I_{-i}$ and ${\tilde I}_{-j}$. To find the matrix
model meaning of these cuts, consider the analytic function: \bea
\label{yu}
\kappa(z)&:=&u_0(z)-u_2(z)=2u_0(z)+u_0(-z)=
2\omega_0(z)+\omega_0(-z)-W'(z)=\nn\\
&=&\int{d\lambda ' \rho(\lambda')\left[\frac{2}{z-\lambda'}-
    \frac{1}{z+\lambda'}\right]}-W'(z)~~, \eea where we used relations
(\ref{shiftomega}) and (\ref{tdef}), which imply: \be
2t(z)+t(-z)=-W'(z)~~.  \ee Similarly, we consider: \bea
\label{wu}
\tau(z)&:=&u_0(z)-u_1(z)=u_0(z)-u_0(-z)=\omega_0(z)-\omega_0(-z)-[W'(z)-W'(-z)]=\nn\\
&=&\int{d\lambda ' \rho(\lambda')\left[\frac{1}{z-\lambda'}+
    \frac{1}{z+\lambda'}\right]}-[W'(z)-W'(-z)]~~, \eea where again we
used (\ref{shiftomega}) and (\ref{tdef}).

\begin{figure}[bt]
\begin{center}
  \scalebox{0.7}{\input{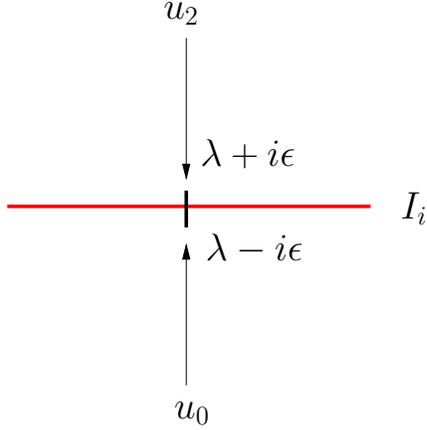}}
\end{center}
\caption{The values of $u_0(\lambda-i0)$ and $u_2(\lambda+i0)$ agree along a
  cut $I_i$, since the latter connects the branches $u_0$ and $u_2$. A
  similar argument holds for the other types of cuts.}
\label{cut_pv}
\end{figure}

The definition of the cuts $I_i$ and ${\tilde I}_j$ implies (see
figure \ref{cut_pv}): \bea u_0(\lambda\pm i 0)&=&u_2(\lambda \mp
i0)~~{\rm for}~~\lambda \in I_i~~
{\rm with~~}i=1\dots d~~\nn\\
u_1(\lambda\pm i 0)&=&u_2(\lambda \mp i0)~~{\rm for}~~\lambda \in
I_{-i}~~
{\rm with~~}i=1\dots d~~\\
u_0(\lambda\pm i 0)&=&u_1(\lambda \mp i0)~~{\rm for}~~\lambda \in
{\tilde I}_j~~{\rm with~~}j=-\delta\dots \delta~~.\nn \eea Therefore,
we have: \bea {\cal P} (\kappa(\la))&=&0~~{\rm for}~~\lambda \in I_i~~
{\rm with~~}i=1\dots d~~\nn\\
{\cal P}(\kappa(-\lambda))&=&0~~{\rm for}~~\lambda \in I_{-i}~~
{\rm with~~}i=1\dots d~~\\
{\cal P}(\tau(\lambda))&=&0~~{\rm for}~~\lambda \in {\tilde I}_j~~
{\rm with~~}j=-\delta\dots \delta~~.\nn \eea

Writing these principal values using (\ref{yu}) and (\ref{wu}), we
find: 
\bea
\int{d\lambda ' \rho(\lambda')\left[\frac{2}{\lambda-\lambda'}-
    \frac{1}{\lambda+\lambda'}\right]}&=&W'(\lambda)~~,~~~~~~~~~~~~~~
\lambda \in I_i\ ,\label{atypecut} \\
\int{d\lambda ' \rho(\lambda')\left[\frac{2}{\lambda+\lambda'}-
    \frac{1}{\lambda-\lambda'}\right]}&=&-W'(-\lambda)~~,~~~~~~~~~
\lambda \in I_{-i}\ ,\label{btypecut}\\
\int{d\lambda ' \rho(\lambda')\left[\frac{1}{\lambda-\lambda'}+
    \frac{1}{\lambda+\lambda'}\right]}&=&W'(\lambda)-W'(-\lambda)~,\ 
\lambda \in {\tilde I}_j\ . \label{ctypecut} \eea Notice that the second
equation is equivalent with the first under the substitution
$\lambda\rightarrow -\lambda$ and therefore contains no new
information.

Assuming that all cuts are disjoint from each other, the analytic
function $u_2$ must be continuous across ${\tilde I}_j$ (since the
cuts of $u_2$ are $I_j$ and $I_{-j}$). Therefore, we have the
following relations for $\lambda\in {\tilde I}_j$: \bea
u_2(\lambda+i0)=u_2(\lambda-i0)&\Longleftrightarrow&
u_0(\lambda+i0)+u_1(\lambda+i0)=u_0(\lambda-i0)+u_1(\lambda-i0)
\Longleftrightarrow \nn\\
u_0(\lambda+i0)-u_0(\lambda-i0)&=&u_1(\lambda-i0)-u_1(\lambda+i0)~~.
\eea Using $u_0(z)=u_1(-z)$ in the right hand side gives: \be
\label{rho_sym}
\rho(\lambda)=\rho(-\lambda)~~,~~\lambda \in {\tilde I}_j~~, \ee where
we also used the identity: \be
\label{rho_omega}
u_0(\lambda-i0)-u_0(\lambda+i0)=2\pi i \rho(\lambda)~~,~~\lambda \in
\R~~.  \ee Thus $\rho$ takes symmetric values along the cuts ${\tilde
  I}_j$ and ${\tilde I}_{-j}$ (in particular, $\rho(\lambda)$ is
symmetric along ${\tilde I}_0$).

Finally, we can identify the filling fractions of the matrix model:
\bea
\label{eq:fillingfractions}
S_i &=& \int_{A_i} \lambda ~~,~~\mathrm{for}~~ i=1\dots d ~,\nn\\
{\tilde S}_j &=& 2\int_{{\tilde A}_j} \lambda~~,~~
\mathrm{for}~~ j=1\dots \delta \\
{\tilde S}_0 &=& \int_{{\tilde A}_0} \lambda~~.\nn \eea The factors
two in the second equation follows from the fact that ${\tilde I}_j$
and ${\tilde I}_{-j}$ support symmetric distributions of eigenvalues
for $j\neq 0$.

{}From the gauge theory point of view, we can interpret
(\ref{atypecut}) as the quantum correction to the classical vacuum
configuration (\ref{rootsii}) and (\ref{ctypecut}) as the correction
to (\ref{rootsi}). The different filling fractions correspond to the
different types of vacua we found in Section 2.1.  In particular, a
fraction ${\tilde S}_j$ with $j\neq 0$ corresponds to a vacuum in
which the final $U(N_j)$ gauge group is embedded diagonally, while
${\tilde S}_0$ corresponds to the vacuum with an orthogonal or
symplectic unbroken gauge group. Of course, up to now we imposed the
strict large $N$ limit and therefore the curve can not distinguish
between the two types of orientifolds.  In section 5 we will argue
that this difference can be accounted for by including a logarithmic
term in the potential $V$ of the model. Let us therefore study this
situation next.

\subsection{Riemann surface in the presence of a logarithmic interaction}

In this subsection, we analyze the Riemann surface (\ref{curve}) in
the presence of a logarithmic term $t_{-1}\ln z$ in the potential $V$.
We shall show that turning on such interactions produces supplementary
cuts, which correspond to eigenvalues accumulating along new loci.
The purpose of the present section is to identify these novel cuts for
small values of $t_{-1}$ --- this will important in our discussion of
the first order loop equations in section \ref{sec:firstorder}.

To understand the generic situation, first note that $p(z)$ and $q(z)$
acquire poles at the origin for $t_{-1}{\neq}0$. To display this, we
decompose $t(z)={\bar t}(z)-\frac{t_{-1}}{z}$, with ${\bar
  t}(z):=\frac{-2W'(z)+W'(-z)}{3}$ a polynomial.  Further, we define
polynomials $\bar p(z)$ and $\bar q(z)$ by substituting $\bar f$,
$\bar g$, $\bar t$ in \erf{pq}. This results in: \bea
\label{pqlog}
p(z) &=& \bar p(z) - \frac{t_{-1}}z\big(\bar t(z){-}\bar t(-z)\big) +
\frac{t_{-1}^{\,2}}{z^2}
\qquad \nn \\
q(z) &=& \bar q(z) + \frac{t_{-1}}z\big( \bar f(z){-}\bar f(-z) +[\bar
t(z){-}\bar t(-z)] [\phi{-}\bar t(z){-}\bar t(-z)]
\big) \\
&& \qquad \qquad - \frac{t_{-1}^{\,2}}{z^2}\big(2\phi - \bar t(z) -
\bar t(-z) \big)\ , \nn \eea where we used \eqref{eq:f-decomp} and
\eqref{eq:gbar-deg} and absorbed the $s$ dependence in $t_{-1}$.  To
write (\ref{curve}) in polynomial form, we multiply the equation by
$z^3$ and define a new variable $y:=zu$: \be
\label{polycurve}
y^3-P(z)y-zQ(z)=(y-y_0(z))(y-y_1(z))(y-y_2(z))=0~~, \ee where
$y_1(z)=-y_0(-z)$, $y_2(z)=-y_0(z)-y_1(z)$ and $P(z)=z^2p(z)$ and
$Q(z):=z^2q(z)$ are even polynomials. This curve admits the symmetry:
\be (z,y)\rightarrow (-z,-y)~~.  \ee We shall view (\ref{polycurve})
as a deformation of the curve: \be
\label{barcurvey}
{\bar y}^3-z^2{\bar p}(z){\bar y} -z^3{\bar q}(z)= ({\bar y}-{\bar
  y}_0(z))({\bar y}-{\bar y}_1(z)) ({\bar y}-{\bar y}_2(z))=0\ , \ee
which is obtained from: \be
\label{barcurve}
u^3-{\bar p}(z)u -{\bar q}(z)=(u-{\bar u}_0(z))(u-{\bar u}_1(z))
(u-{\bar u}_2(z))=0\ , ~~ \ee where ${\bar u}_2(z)=-{\bar
  u}_0(z)-{\bar u}_1(z)$ and ${\bar u}_1(z)={\bar u}_0(-z)$, by
performing the birational transformation ${\bar y}=zu$.

\begin{figure}[hbtp]
\begin{center}
  \scalebox{0.4}{\input{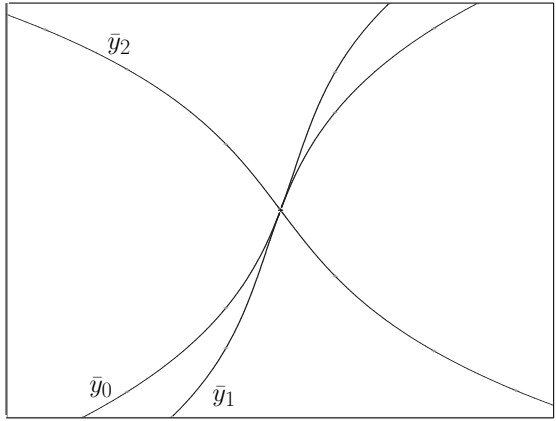}}
\end{center}
\caption{Branching of the curve (\ref{polycurve}) at the origin for 
  $t_{-1}=0$.}
\label{deg_branches}
\end{figure}

We are interested in the generic behavior of (\ref{polycurve}) as
$t_{-1}$ approaches zero.  In this limit, the curve degenerates to
(\ref{barcurvey}), whose first two branches ${\bar y}_0$ and ${\bar
  y}_1$ are tangent in a cusp at ${\bar y}=z=0$, where they also meet
the third branch ${\bar y}_2$ (figure \ref{deg_branches}).

\begin{figure}[hbtp]
\begin{center}
  \scalebox{0.35}{\input{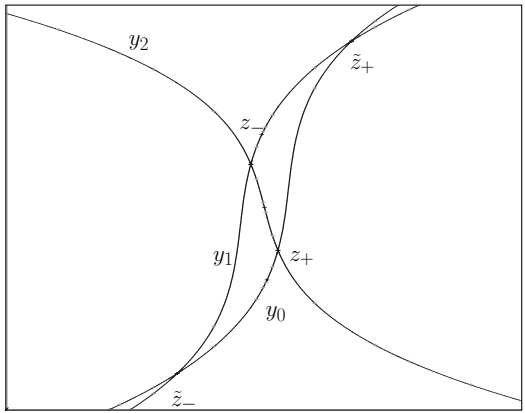}}
\end{center}
\caption{Deformed curve after addition of a logarithmic term in the matrix 
  model potential, in the birationally equivalent coordinates $(y,z)$.
  }
\label{def_curve}
\end{figure}

The geometry of (\ref{polycurve}) for a small value of $t_{-1}$ is
described in figure \ref{def_curve}.  The logarithmic deformation of
the model generates four new branch points of the birationally
transformed curve, which are indexed by $z_\pm$ and ${\tilde z}_\pm$.
These are symmetric with respect to the origin, namely $z_-=-z_+$ and
${\tilde z}_-=-{\tilde z}_+$.

From this geometry, one finds that the cut ${\tilde I}_0$ splits
into two new cuts ${\tilde I}_0^+=[{\tilde z}_+,{\tilde b}_0]$, and
${\tilde I}_0^-=[{\tilde a}_0, {\tilde z}_-]$, which are distributed
symmetrically with respect to the origin. Both of these cuts connect
the branches $0$ and $1$.  One also finds two new cuts $I_0=(-\infty,
z_+)$ and $-I_0=(z_-,+\infty)$, which connect the pairs of branches
$(0,2)$ and $(1,2)$ respectively.  As in \cite{Cachazo:2001jy}, we
shall `regularize' our curve by choosing a large $\Lambda>0$ and
replacing these cuts with $I_0=(-\Lambda, z_+)$ and
$-I_0=(z_-,+\Lambda)$.

\begin{figure}[hbtp]
\begin{center}
  \scalebox{0.5}{\input{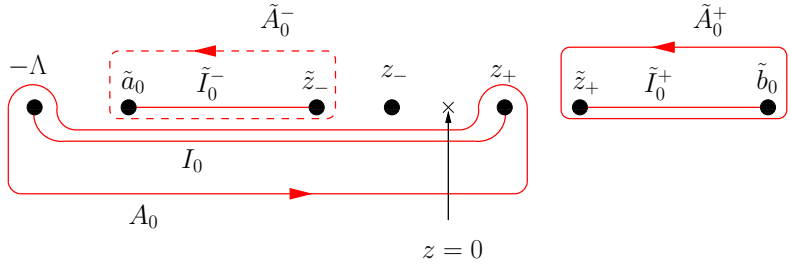}}
\end{center}
\caption{The new cuts of $u_0$
  generated by a logarithmic term in the potential (viewed as cuts of
  the birationally transformed curve (\ref{polycurve})). The cut
  $-I_0$ corresponds to branching of $u_1$ with $u_2$, and
  consequently it is not shown.  The figure also indicates the new cycles
  $A_0$, ${\tilde A}_0^\pm$ on the Riemann surface.  Solid cycles lie
  on the sheet $u_0(z)$, while the dashed cycle lies on $u_1(z)$.  The curves
  $\gamma_0$,${\tilde \gamma}_0^\pm$ mentioned in the text 
  are the projections of $A_0$ and 
  ${\tilde A}_0^\pm$ onto the $z$-plane. }
\label{newcuts}
\end{figure}

These cuts produce new A-type cycles $A_0$, ${\tilde A}_0^{\pm}$ (and
associated B-cycles), which project onto the $z$-plane to the curves
$\gamma_0, {\tilde \gamma}^\pm_0$, see figure \ref{newcuts}.  The
homology classes ${\tilde A}_0^\pm$ are interchanged by the symmetry
$z\rightarrow -z$, which also interchanges the homology class of $A_0$
with that of $-A_0$.

The cuts $I_0$ and ${\tilde I}^{\pm}_0$ give the A-type periods: \bea
\label{extra_periods}
\oint_{\gamma_0}{\frac{dz}{2\pi i}\omega_0(z)}&=&
\oint_{A_0}{\frac{dz}{2\pi i}\omega(z)}~~,\nn\\
\oint_{{\tilde \gamma}^\pm_0}{\frac{dz}{2\pi i}\omega_0(z)}&=&
\oint_{{\tilde A}^\pm_0}{\frac{dz}{2\pi i}\omega(z)}~~.\nn\\
\eea The cut $-I_0$ gives the opposite of the $A_0$-period. The $\Z_2$
symmetry also implies: \be \oint_{{\tilde
    \gamma}_0^{-}}{\frac{dz}{2\pi i}\omega_0(z)}= \oint_{{\tilde
    \gamma}_0^{+}}{\frac{dz}{2\pi i}\omega_0(z)}~~.  \ee

\ 

The presence of new cuts for the log-deformed model shows that
eigenvalues can accumulate along $I_0$ and ${\tilde I}_0^{\pm}$ in the
large $N$ limit (no eigenvalues accumulate on $-I_0$ since this is not
a cut of $\omega_0$).  Accordingly, the microcanonical ensemble of the
log-deformed model involves the supplementary constraint:

\bea
\label{S_0constraint}
\oint_{\gamma_0}{\frac{dz}{2\pi i}\langle \omega(z)\rangle}&=&S_0~~,
\eea while the constraint $\oint_{{\tilde \gamma}_0}{\frac{dz}{2\pi
    i}\langle \omega(z)\rangle}= {\tilde S}_0$ of the undeformed model
is replaced by: \be \oint_{{\tilde \gamma}^+_0}{\frac{dz}{2\pi
    i}\langle \omega(z)\rangle}= \frac{1}{2}{\tilde S}_0~~.  \ee

This becomes part of conditions (\ref{fconstraint}) for the case
$t_{-1}\neq 0$.  For a consistent limit as $t_{-1}\rightarrow 0$, one
takes $S_0$ to be a function of $t_{-1}$ which tends to zero in that
limit. Note also that with the additional constraint
(\ref{S_0constraint}) we have to fix $d+\delta+2$ periods in the
log-deformed model, in agreement with the parameter count in
(\ref{eq:f-gev-coeff}).

\section{Chemical potentials at large $N$}

In this section, we study the planar limit of our models for the case
$t_{-1}=0$.  Our main purpose is to give a {\em proof} \footnote{The
  original paper \cite{DV1} gives a beautiful intuitive argument for
  the existence of such a relation, without providing a rigorous proof
  (what is missing is to show that the Riemann surface B-periods
  indeed equal appropriately defined chemical potentials). A
  derivation of this relation was later given in \cite{Whitham,
    Whitham2, Whitham3} (using older results of \cite{Jurk}), though a
  clear construction of the microcanonical ensemble (which is implicit
  in that argument) was not given there. In the present paper, we are
  dealing with the more complicated case of a non-hyperelliptic
  Riemann surface, which underscores the usefulness of having a clear
  proof of such relations.} of an appropriate set of special geometry
relations. The path we shall follow is based on our construction of
the microcanonical ensemble, combined with a modification of an
argument due to \cite{Whitham}. Namely, we shall show that the
chemical potentials $\mu$ reduce in the large $N$ limit to certain
B-type periods $\Pi$ of our Riemann surface. Then special geometry
conditions of the type found in \cite{DV1} follow from the standard
equation (\ref{quantum_periods}) which expresses the chemical
potentials as derivatives of the {\em microcanonical} partition
function with respect to the filling fractions. In particular,
equation (\ref{quantum_periods}) is the appropriate finite $N$
generalization of the special geometry constraints.  We also show that
Whitham-type relations of the type found in \cite{Whitham} arise as
the planar limit of certain finite $N$ equations which follow
naturally in the microcanonical ensemble.  For simplicity, we assume
$t_{-1}=0$ for most of the present section.  The generalization to the
case $t_{-1}\neq 0$ is entirely obvious, but notationally tedious.

Let us hence assume $t_{-1}=0$.  Since $\rho(\lambda)$ in our models
develops a nonzero value along the cuts $I_i$ and ${\tilde I}_j$ with
$i=1\dots d$ and $j=-\delta \dots \delta$, we have $d+2\delta+1$
filling fractions.  To construct the microcanonical ensemble, we pick
intervals $\Delta_i$ ($i=1\dots d$) and ${\tilde \Delta}_j$
($j=-\delta\dots \delta$), such that ${\tilde \Delta}_{-j}=-{\tilde
  \Delta}_j$ for all $j$.  In the large $N$ limit, we shall assume
that the intervals $\Delta_i$ and ${\tilde \Delta}_j$ contain the cuts
$I_i$ and ${\tilde I}_j$.  The symmetry property (\ref{rho_sym}) shows
that the filling fractions of ${\tilde I}_j$ and ${\tilde I}_{-j}$ are
equal, and we shall take ${\tilde S}_0 \dots {\tilde S}_\delta$ to be
the independent quantities: \bea
\label{csrt}
\int_{I_i}{d\lambda \rho_0(\lambda)}&=&S_i~~
{\rm for}~~i=1\dots d\nn\\
\int_{{\tilde I}_j}{d\lambda \rho_0(\lambda)}&=&\frac{1}{2}{\tilde
  S}_j
~~{\rm for}~~j=1\dots \delta~~\\
\int_{{\tilde I}_0}{d\lambda \rho_0(\lambda)}&=&{\tilde S}_0~~.\nn
\eea

Then the $d+\delta+1$ constraints (\ref{csrt}) completely fix the
$d+\delta+1$ deformations of (\ref{curve}) encoded by the polynomials
$f$ and $g_{ev}$. In the (grand) canonical ensemble, this is
implemented by introducing chemical potentials $\mu_i$ and ${\tilde
  \mu}_j$ with $i=1\dots d$ and $j=0\dots \delta$.

The planar (grand) canonical generating function reads: \bea {\cal
  F}_0(t,\mu) &=& \int{d\lambda W(\lambda)\rho_0(\lambda)}-\frac{1}{2}
\int{d\lambda}\int{d\lambda' {\cal K}(\lambda,\lambda')\rho_0(\lambda)
  \rho_0(\lambda')} \nn\\
&&\qquad +\sum_{i=1}^d{\mu_iS_i} +\sum_{j=0}^\delta{{\tilde
    \mu}_j{\tilde S}_j} \eea where \be {\cal
  K}(\lambda,\lambda')=2\ln|\lambda-\lambda'|-\ln|\lambda+\lambda'|~~.
\ee

This gives the planar limit of the microcanonical generating function:
\be
\label{F_0}
F_0(t,S)=\frac{1}{2}\int{d\lambda}\int{d\lambda'{\cal
    K}(\lambda,\lambda')
  \rho_0(\lambda)\rho_0(\lambda')}-\int{d\lambda
  W(\lambda)\rho_0(\lambda)}~~, \ee with the constraints (\ref{csrt}).

{\bf Observation} Remembering relation (\ref{rho_sym}), one finds the
functional derivatives: \bea
\label{F_rho}
\frac{\delta F_0}{\delta \rho(\lambda)}&=& \int{d\lambda' {\cal
    K}(\lambda,\lambda')\rho_0(\lambda')}-W(\lambda)
~~,~~\lambda\in I:=\cup_{i=1\dots d}{I_i}\nn\\
\frac{\delta F_0}{\delta \rho(\lambda)}&=& \int{d\lambda' {\tilde
    {\cal K}}(\lambda,\lambda')\rho_0(\lambda')}-
W(\lambda)-W(-\lambda) ~~,~~\lambda\in {\tilde I}:=\cup_{j=-\delta
  \dots \delta}{{\tilde I}_j}~~, \eea where: \be {\tilde {\cal
    K}}(\lambda,\lambda')= {\cal K}(\lambda, \lambda')+{\cal
  K}(-\lambda,\lambda')=
\ln|\lambda+\lambda'|+\ln|\lambda-\lambda'|~~.  \ee Note that equating
(\ref{F_rho}) to constants and further differentiating with respect to
$\lambda$ gives the planar equations of motion 
(\ref{atypecut},\ref{btypecut},\ref{ctypecut}). This
observation can be viewed as an intuitive justification for the
rigorous procedure discussed below.

\subsection{The primitive of $\kappa$ along the real axis}

To extract the large $N$ chemical potentials, we shall be interested
in the `restriction' of the function $\kappa$ of subsection
\ref{sec:phys} along the real axis, which we define by: \be
\kappa_p(\lambda)=\frac{1}{2}\left[\kappa(\lambda+i0)+\kappa(\lambda-i0)\right]
\ee for any real $\lambda$.  If $\lambda$ is a real value lying
outside the union of $I_i$, then $\kappa_p(\lambda)$ equals
$\kappa(\lambda)$, the quantity obtained by substituting $\lambda$ for
$z$ in (\ref{yu}). Taking the principal value of (\ref{yu}) along the
real axis gives: \be \kappa_p(\lambda):=\int{d\lambda'
  \rho_0(\lambda')K(\lambda,\lambda')}-W'(\lambda)~~, \ee where
$K(z,z'):=\frac{2}{z-z'}-\frac{1}{z+z'}$.  Consider now the function
$\phi:\R\rightarrow \C$ defined through: \be
\label{phi_def}
\phi(\lambda):=\int{d\lambda'{\cal K}(\lambda,\lambda')
  \rho_0(\lambda')}-W(\lambda)~~.
\end{equation}
Noticing that $K(\lambda,\lambda')=\frac{\partial }{\partial
  \lambda}{\cal K}(\lambda,\lambda')$ shows that $\phi$ is a primitive
of $\kappa_p$: \be
\label{dphi}
\frac{d\phi}{d\lambda}=\kappa_p~~.
\end{equation}

As discussed in the previous section, the equations of motion
(\ref{eq:EOM}) amount to the requirement that $\kappa_p$ vanishes
along each of the intervals $I_i$: \be \kappa_p(\lambda)=0~~{\rm
  for}~~\lambda \in I=\cup_{i=1}^d{I_i}~~.  \ee This means that $\phi$
is constant along each of these intervals: \be
\label{constancy}
\phi(\lambda)=\xi_i={\rm~constant~for~}\lambda\in I_i~~.  \ee Then the
jump in the value of $\kappa_p$ between consecutive cuts can be
obtained by integrating (\ref{dphi}) between their endpoints: \bea
\label{xi_int0}
\xi_{i+1}-\xi_i &=& \int_{b_i}^{a_{i+1}}{d\lambda \kappa(\lambda)}~~.
\eea

\subsection{The primitive of $\tau$ along the real axis}

An entirely similar discussion can be given for the function
$\tau(z)$, whose `restriction' to the real axis is: \be
\tau_p(\lambda)=\frac{1}{2}\left[\tau(\lambda+i0)+\tau(\lambda-i0)\right]=
\int{d\lambda' \rho_0(\lambda'){\tilde K}(\lambda,\lambda')}-
[W'(z)-W'(z')]~~, \ee where ${\tilde
  K}(z,z'):=\frac{1}{z-z'}-\frac{1}{z+z'}$.  Its primitive \be
\label{psi_def}
\psi(\lambda):=\int{d\lambda'{\tilde {\cal K}}(\lambda,\lambda')
  \rho_0(\lambda')}-W(\lambda)-W(\lambda')~~ \ee can be integrated in
between the cuts ${\tilde I}_j$ to give: \bea
\label{xi_int0more}
{\tilde \xi}_{j+1}-{\tilde \xi}_j &=& \int_{{\tilde b}_j}^{{\tilde
    a}_{j+1}}{d\lambda \tau(\lambda)}~~, \eea where ${\tilde \xi}_j$
is the constant value of $\psi$ along ${\tilde I}_j$ (as required by
the second planar equation of motion): \be \tau_p(\lambda)=0~~{\rm
  for}~~\lambda\in {\tilde I}= \cup_{j=-\delta\dots \delta}{{\tilde
    I}_j}~~.  \ee

\subsection{The large $N$ chemical potentials}

Differentiating (\ref{F_0}) with respect to $S_i$ for some $i<d$ and
using relation (\ref{phi_def}) gives: \be
\label{mu_xi}
\mu^{(0)}_i-\mu_d^{(0)}=\frac{\partial}{\partial S_i}F_0(t,S)=
\int_{I\cup {\tilde I}} {d\lambda \frac{\partial
    \rho_0(\lambda)}{\partial S_i}\phi(\lambda)}= \xi_i-\xi_d~~,
\end{equation}
where we took the dependent filling fraction to be $S_d$.

To arrive at the last equality, we used equation (\ref{constancy}) and
the constraints (\ref{csrt}) and (\ref{Snorm}).  Relation (\ref{mu_xi})
shows that the chemical potentials associated with $I_i$ coincide with
the quantities $\xi_i$ in the planar limit, up to a common additive
constant. Using relation (\ref{xi_int0}), we obtain: \be
\label{xi_int1}
\mu^{(0)}_{i+1}-\mu^{(0)}_i=\int_{b_i}^{a_{i+1}}{d\lambda
  \kappa(\lambda)} \ee

A similar argument for the differential with respect to ${\tilde S}_j$
(using relation (\ref{psi_def}) and the last two constraints in
(\ref{csrt})) gives ${\tilde \mu}_i^{(0)}={\tilde \xi}_i$ and: \be
\label{xi_int_more}
{\tilde \mu}^{(0)}_{j+1}-{\tilde \mu}^{(0)}_j= \int_{{\tilde
    b}_j}^{{\tilde a}_{j+1}}{d\lambda \tau(\lambda)}~~.  \ee

\subsection{Geometric expression for the large $N$ chemical potentials}

\begin{figure}[hbtp]
\begin{center}
  \scalebox{0.6}{\input{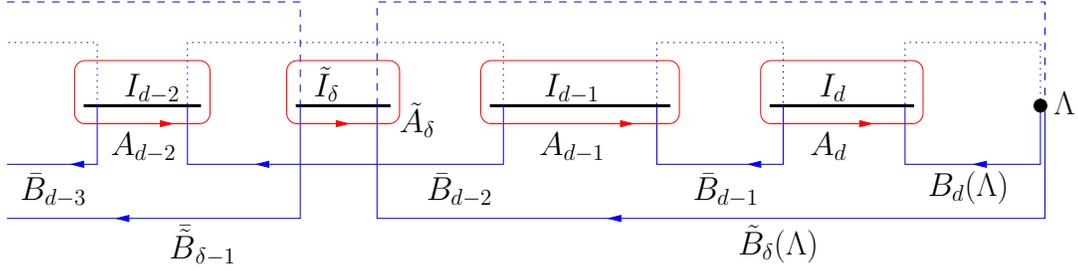}}
\end{center}
\caption{Choice of B-cycles on the large $N$ Riemann surface.
  We only indicate a few cuts close to the point $\Lambda$.  The
  cycles ${\bar B}_i,{\bar {\tilde B}}_j$ and $B_d(\Lambda), {\tilde
    B}_\delta(\Lambda)$ are defined such that, when crossing a cut
  going upwards along these cycles, one moves from the branch $u_0$ to
  the branch $u_1$ (drawn dashed) or $u_2$ (dotted) respectively.
  With the orientation of the Riemann surface induced by its complex
  structure, and with the indexing $\alpha$ of cuts explained the
  text, this implies the intersections $A_\alpha \cap {\bar
    B}_\alpha=+1$ for the re-indexed cycles (note that the cycles
  $A_\alpha$ and ${\bar B}_\alpha$ intersect in a single point, which
  lies on the branch $u_0$).}
\label{Bcycles}
\end{figure}

Together with (\ref{yu}), equation (\ref{xi_int1}) gives: \be
\label{xi_int}
\mu^{(0)}_{i}-\mu^{(0)}_{i+1}= \int_{b_i}^{a_{i+1}}{dz
  [u_2(z)-u_0(z)]} =\oint_{{\bar B}_i}{dz u(z)}~~,
\end{equation}
where ${\bar B}_i$ are cycles on the large $N$ Riemann surface chosen
as explained in figure \ref{Bcycles}. Similarly, we find: \be
\label{xi_int_less}
{\tilde \mu}^{(0)}_{j}-{\tilde \mu}^{(0)}_{j+1}= \int_{{\tilde
    b}_j}^{{\tilde a}_{j+1}}{dz [u_1(z)-u_0(z)]} =\oint_{{\tilde {\bar
      B}}_j}{dz u(z)}~~, \ee with ${\tilde {\bar B}_j}$ chosen as in
figure \ref{Bcycles}.

\ 

Consider the cycles ${\hat B}_i=\sum_{k=i}^{d-1}{\bar B}_k$ for all
$i=1\dots d-1$. Then (\ref{xi_int}) implies: 
\be
\label{foo}
\mu^{(0)}_i=\mu^{(0)}_d+ \oint_{{\hat B}_i}{dz
  u(z)}~~{\rm~for~}~~i=1\dots d-1~~.
\end{equation}
The quantity $\mu_d^{(0)}$ is undetermined and can be fixed
arbitrarily.  Following \cite{DV1}, we take
$\mu_d^{(0)}=\oint_{B_d}{dz u(z)}$, where $\Lambda$ is a point close
to $+\infty$ and $B_d(\Lambda)$ is a path shown in figure
\ref{Bcycles}. Defining $B_i(\Lambda)={\hat B}_i+B_d(\Lambda)$ for all
$i=1\dots d-1$, equation  (\ref{foo}) gives:
\be
\label{mu_Pi}
\mu^{(0)}_i=\Pi_i {\rm~~for~~}i=1\dots d~~,
\ee
with: \be
\label{Pi}
\Pi_i:=\int_{B_i}{dz u(z)}~~.
\end{equation}

A similar construction for the cuts ${\tilde I}_j$ gives: \be
\label{mu_Pi_more}
{\tilde \mu}^{(0)}_j={\tilde \Pi}_j {\rm~~for~~}j=0\dots \delta~~, \ee
where: \be
\label{Pi_more}
{\tilde \Pi}_j:=\int_{{\tilde B}_j}{dz u(z)}~~, \ee with cycles
${\tilde B}_j$ chosen analogously (figure \ref{Bcycles}): \be {\tilde
  B}_j={\tilde B}_\delta (\Lambda)+ \sum_{k=j}^{\delta-1}{{\tilde
    {\bar B}}_k}~~.  \ee Relations (\ref{mu_Pi}) and
(\ref{mu_Pi_more}) show that the chemical potentials $\mu$ are the
finite $N$ analogues of the periods $\Pi$.

Also remember that the filling fractions can be expressed as periods
of $udz$ over the cycles $A_\alpha$ of figure \ref{Acycles}: \be
S_\alpha=\oint_{\gamma_\alpha}{\frac{dz}{2\pi i}
  \omega_0(z)}=\oint_{\gamma_\alpha}{\frac{dz}{2\pi i}u_0(z)}=
\oint_{A_\alpha}{\frac{dz}{2\pi i}u(z)}~~.
\end{equation}
In the second equality, we used relation (\ref{shiftomega}) and the
fact that $t(z)$ is a polynomial.

Defining $q:=d+\delta$, let us index the cuts by $\alpha=-q\dots q$ in the
order in which the intervals $I_i$, ${\tilde I}_j$ appear from left to
right. We also let ${\cal A}_\alpha$ and ${\bar {\cal B}}_\alpha$ respectively
${\cal B}_\alpha$ be the corresponding cycles $A_i, {\tilde A}_j$ and ${\bar
B}_i, {\tilde {\bar B}}_j$ respectively $B_i$, ${\tilde B}_j$ indexed in this
order.  It is then clear from figure \ref{Bcycles} that ${\cal A}_\alpha\cap
{\bar {\cal B}}_\alpha= -{\cal A}_\alpha\cap {\bar {\cal B}}_{\alpha-1}=+1$
and ${\cal A}_\alpha\cap {\bar {\cal B}}_\beta=0$ if $\beta\neq \alpha,
\alpha-1$. This gives ${\cal A}_\alpha\cap {\cal B}_\beta=-{\cal B}_\beta\cap
{\cal A}_\alpha=\delta_{\alpha\beta}$.  Since we also have ${\cal
A}_\alpha\cap {\cal A}_\beta= {\cal B}_\alpha\cap {\cal B}_\beta=0$, it
follows that ${\cal A}_\alpha,{\cal B}_\beta$ have canonical intersection
form.  Thus we find a canonical system of cycles ${\cal A}_\alpha, {\cal
B}_\alpha$ with $\alpha= -q\dots q$ : 
\be 
{\cal A}_\alpha\cap {\cal B}_\beta=- {\cal B}_\alpha\cap {\cal
  A}_\beta=\delta_{\alpha,\beta}~~,~~ {\cal A}_\alpha\cap {\cal A}_\beta=
  {\cal B}_\alpha\cap {\cal B}_\beta=0 ~~{\rm for~all}~~ \alpha=-q\dots q~~.
\ee

\subsection{Whitham-type relations}

Let us now show how a finite $N$ version of a set of Whitham-type
constraints (similar to those found in \cite{Whitham} for the case of
one-matrix models) can be extracted from the microcanonical ensemble.
For the rest of this subsection, we shall allow for a logarithmic term
in the potential--- as we shall see in a moment, this has interesting
effects on our Whitham-type relations.

Differentiating (\ref{F_def}) with respect to $t_m$ gives: \be
\label{l_def}
\frac{\partial}{\partial t_m}F(t,S) =-\frac{\partial}{\partial
  t_m}{\cal F}(t,\mu)|_{\mu_\alpha=\mu_\alpha(t,S)} =l_m~~, \ee where
we defined the averaged loop operators: $l_m:=-\frac{1}{N}\langle tr
\partial_{t_m} V(M)\rangle= -\frac{1}{N(m+1)}\langle
tr(M^{m+1})\rangle$ for $m>0$ and $l_{-1}:=-\frac{1}{N}\langle tr(log
M)\rangle$ for $m=-1$.  Equations (\ref{l_def}) supplement relations
(\ref{quantum_periods}).  We have: \be
\label{l_m}
-l_m=
\begin{cases}
  \frac{1}{m+1}\int{d\lambda \langle \rho(\lambda)\rangle
    \lambda^{m+1}}= \frac{1}{m+1}\oint_{\gamma} {\frac{dz}{2\pi
      i}{z^{m+1}\langle \omega(z)\rangle }}
  ~{\rm~~if~~} m>0\\
  \int{d\lambda \langle \rho(\lambda)\rangle \log \lambda}=
  \oint_{\gamma}{\frac{dz}{2\pi i}{\langle \omega(z)\rangle \ln z}}
  ~{\rm~~if~~} m=-1
\end{cases}~~,
\ee
where $\gamma$ is a counterclockwise contour that encircles all 
eigenvalues.  Combined with (\ref{l_m}), equations (\ref{l_def}) give a
finite $N$ version of a set of Whitham-like constraints; these reduce
to standard Whitham conditions in the planar limit and for the case
$t_{-1}=0$.

\subsubsection{Geometric form of the large $N$ Whitham-type constraints}

For $m>0$, one can deform the contour $\gamma$ toward infinity, thus
picking up contributions from $z=\infty$. This gives: \be
\label{l_as}
l_m=\frac{1}{m+1}Res_{\zeta=0} \left[\frac{\langle
    \omega(1/\zeta)\rangle }{\zeta^{m+3}}\right]~~{\rm for}~~m>0~~,
\ee
where $\zeta=1/z$.

In the large $N$ limit, one has $\langle
\omega(z)\rangle=\omega_0(z)$, with $\omega_0(z)$ a branch of the
Riemann surface (\ref{curve}). In this case, equations (\ref{l_as})
and the last equation in (\ref{l_m}) determine the derivatives
$\frac{\partial F_0}{\partial t_m}$ in terms of the coefficients of
$V$ and $f,g$; these are Whitham-type conditions analogous to those of
\cite{Whitham}.  As shown above, the derivatives $\frac{\partial
  F_0}{\partial S_\alpha}$ are also determined by the Riemann surface.
Together with the Whitham relations, this allows one to determine the
planar generating function $F_0(t,S)$ up to an additive
constant\footnote{We stress that equations (\ref{quantum_periods}) only
  determine $F_0(t,S)$ up to the addition of an arbitrary function of
  $t_m$. One needs the Whitham constraints in order to fix the
  $t$-dependence of $F_0$.}, given the planar limits 
$l^{(0)}(t,S)$ and $\Pi(t,S)$.

\section{First order analysis of the loop equations}\label{sec:firstorder}

Consider the first order large $N$ expansion of the microcanonical
generating function $F(t,S)$ of the log-deformed model: \be F(t, S) \;
= \; F_0(t,S) \;+\; \tfrac{1}{N} \; F_1(t,S) \;+\; O(N^{-2}) \;.  \ee
The aim of this section is to show that the contribution $F_1$ can be
obtained by differentiation with respect to the parameter $t_{-1}$ in
the potential \erf{eq:potential}: \be F_1(t,S) \;=\; \frac{s}{2} \;
\frac{\partial F_0(t,S)}{\partial t_{-1}}+\Psi(t_{-1},S_0)~~.  \ee

\subsection{The resolvent to order $1/N$}
To establish this, we first note that the multi-point correlators of
the resolvent $\omega$ differ from the corresponding product of one
point functions only at order $1/N^2$: \be \langle \om(z)^2 \rangle =
\langle \omega(z)\rangle^2 + O(1/N^2) \quad , \qquad \langle
\om(z)^2 \om(-z) \rangle = \langle \omega (z)\rangle^2 \langle \omega
(-z) \rangle + O(1/N^2)
  \label{eq:1st-order-corr}
  \ee and so forth. Consider the $1/N$--expansion: \be \langle \omega
  (z) \rangle = \om_0(z) + \frac{1}{N} \omega_1(z) + {\cal
    O}(1/N^2)~~.  \ee 
  Inserting this and \erf{eq:1st-order-corr} into
  the integral formulation (\ref{eq:intthirdloop}) of the loop
  equations we obtain the planar relations: \bea
  \label{lplanar}
  \oint_{\gamma}{\frac{dx}{2\pi i}\frac{2xV'(x)}{z^2-x^2}
    \omega_0(x)}~~~~~~~~~~&~=~&
  \omega_0(z)^2+\omega_0(z)\omega_0(-z)+\omega_0(-z)^2~~,\nn\\
  \oint_{\gamma}{\frac{dx}{2\pi i} \frac{2xV'(x)}{z^2-x^2}
    \omega_0(x)\omega_0(-x)}&~=~&
  \omega_0(z)^2\omega_0(-z)+\omega_0(-z)^2\omega_0(z) \eea 
  and the $O(1/N)$ constraints: 
\bea
\label{lnext}
&&\oint_{\gamma}{\frac{dx}{2\pi i}\frac{2xV'(x)}{z^2-x^2}
  \omega_1(x)}+ s\oint_{\gamma}{\frac{dx}{2\pi i}
  \frac{\omega_0(x)}{z^2-x^2}}
\nn\\[4pt]
&& \qquad \qquad = [2\omega_0(z)+\omega_0(-z)]\omega_1(z)
+~[2\omega_0(-z)+\omega_0(z)]\omega_1(-z)
~~,\\[8pt]
&&\oint_{\gamma}{\frac{dx}{2\pi i}\frac{2xV'(x)}{z^2-x^2}
  [\omega_0(-x)\omega_1(x)+\omega_0(x)\omega_1(-x)]} +
s\oint_{\gamma}{\frac{dx}{2\pi i}
  \frac{\omega_0(x)\omega_0(-x)}{z^2-x^2}}
\nn\\[4pt]
&& \qquad \qquad =
[2\omega_0(z)\omega_0(-z)+\omega_0(-z)^2]\omega_1(z)+
[2\omega_0(-z)\omega_0(z)+\omega_0(z)^2]\omega_1(-z) ~~.\nn 
\eea 
On
the other hand, the filling fraction conditions give: \bea
\oint_{\gamma_i}{\frac{dz}{2\pi i}\,\omega_0}=S_i~~(i=0\dots d)~~,~~
\oint_{{\tilde \gamma}_j}{\frac{dz}{2\pi
    i}\omega_0}=\frac{1}{2}{\tilde S}_j~ ~(j=1\dots
\delta)~~,~~\oint_{{\tilde \gamma}^+_0} {\frac{dz}{2\pi
    i}\omega_0}=\frac{1}{2}{\tilde S}_0~~,~~
  \label{fc0}~~~~~
  \eea and \be \oint_{\gamma_i}{\frac{dz}{2\pi
      i}\,\omega_1}=0~~(i=0\dots d)~~, ~~ \oint_{{\tilde
      \gamma}_j}{\frac{dz}{2\pi i}\,\omega_1}=0~~ (j=1\dots
  \delta)~~,~~ \oint_{{\tilde \gamma}^+_0}{\frac{dz}{2\pi
      i}\,\omega_1}=0~~.
  \label{fc1}
  \ee These include the constraints for the log-deformed model arising
  for the cuts shown in figure \ref{newcuts}.

  Suppose that $\om_0(z)$ is a solution to \erf{lplanar} and \erf{fc0}
  for the potential \erf{eq:potential}.  By differentiating the planar
  loop equations \erf{lplanar} as well as \erf{fc0} with respect to
  $t_{-1}$ one can check that $\omega_1^p:=\frac{s}{2}\frac{\partial
    \omega_0}{\partial t_{-1}}$ is a particular solution of both
  (\ref{lnext}) and (\ref{fc1}).  To show that this is the unique
  solution, it suffices to check that the homogeneous system obtained
  from (\ref{lnext}) by dropping the terms
  $s\oint_{\gamma}{\frac{dx}{2\pi i}\frac{\omega_0(x)}{z^2-x^2}}$ and
  $s\oint_{\gamma}{\frac{dx}{2\pi
      i}\frac{\omega_0(x)\omega_0(-x)}{z^2-x^2}}$ from the two left
  hand sides admits only the trivial solution when supplemented by the
  constraints (\ref{fc1}). To establish this, notice that the
  homogeneous system associated with (\ref{lnext}) results from the
  planar equations (\ref{lplanar}) if one performs infinitesimal
  variations of $\omega_1=\delta \omega_0$, such that the variations are
  independent of the coefficients $t_{-1}\dots t_{d}$ of the potential
  (\ref{eq:potential}). Such variations of $\omega_0$ arise by
  changing the filling fraction parameters $S_\alpha$, which give a
  parameterization of that part of the moduli space of our Riemann
  surface which results by varying the quantities $f,g$ of Subsection
  \ref{subs:le}. Therefore, the solution space to the homogeneous system
  associated with (\ref{lnext}) is the tangent space ${\cal T}$ to
  this part of the moduli space of our surface. Since the
  corresponding moduli space is parameterized by $S_\alpha$, the
  dimension of this solution space equals $d+\delta+2$. On the other
  hand, relations (\ref{fc1}) give $d+\delta+2$ linearly independent
  equations for such solutions. 
  These equations select the vanishing solution of the
  homogeneous system associated with (\ref{lnext}).  Therefore, the
  homogeneous system obtained from (\ref{lnext}) has no nontrivial
  solutions when supplemented by the constraints (\ref{fc1}).  This
  shows that the solution of (\ref{lnext}) and (\ref{fc1}) is uniquely
  determined and equals $\omega_1^p$.
  
  We conclude that the solution of the loop equations has the form:
  \be \langle
  \omega(z)\rangle=\omega_0(z)+\frac{1}{N}\omega_1(z)+{\cal
    O}(1/N^2)~~, \ee where: \be
\label{rel}
\omega_1(z)=\frac{s}{2}\frac{\partial \omega_0(z)}{\partial t_{-1}}~~.
\ee

\subsection{The $1/N$ correction to the microcanonical generating function}
Next we show that a similar relation holds for the $1/N$-expansion of
$F$.  To simplify notation we label all filling fractions $S$, $\tilde
S$ by $S$, keeping the convention that $S_0$ denotes the filling
fraction of $I_0$ in figure \ref{newcuts}.  Equation (\ref{rel}) was
derived under the assumption that the nonsingular part $W$ of our
potential contains a finite number of terms $d$.  Since (\ref{rel})
must hold for any value of $d$, it is clear that it also holds if we
formally allow $W$ to contain an infinity of terms. Therefore, let us
now take $W(z)=\sum_{m=0}^{\infty}{\frac{t_m}{m+1}z^{m+1}}$.  Defining
the loop insertion operator $\frac{d}{dW(z)}=\sum_{m=0}^\infty
{\frac{m+1}{z^{m+1}}\frac{\partial}{\partial t_m}}$ (see e.g.
\cite{Ambjorn:1992gw}), we then have $(m+1)\frac{\partial F}{\partial
  t_m}=\frac{1}{N}\langle tr M^{m+1} \rangle$, for all $m\geq 0$,
which gives the standard relation: \be \langle
\omega(z)\rangle=-\frac{d}{dW(z)}F~~ \ee upon expanding
$\omega(z)=\frac{1}{N}\langle \tr\frac{1}{z-M}\rangle=
\frac{1}{N}\sum_{m=0}^\infty{\frac{1}{z^{m+1}} \frac{\langle \tr
    M^{m+1}\rangle }{m+1}}$.  Here $F$ is the microcanonical
generating function.  Expanding $F=\sum_{j\geq 0}{\frac{1}{N^j}F_j}$,
this gives $\omega_j=-\frac{d}{dW(z)}F_j$. Therefore, relation
(\ref{rel}) implies: \be
\label{F_10}
F_1(t_{-1},\dots t_\infty,S_0\dots S_\infty)=
\frac{s}{2}\frac{\partial F_0(t_{-1}\dots t_\infty, S_0\dots
  S_\infty)}{\partial t_{-1}}+ \Psi(t_{-1},S_0\dots S_\infty)~~, \ee
where $\Psi$ is a {\em universal} function of $t_{-1},S_0\dots
S_\infty$ which does not depend on $(t_j)_{j\geq 0}$. Since this
function is completely independent of $t_0\dots t_\infty$, it knows
nothing about which of these coefficients are zero. To determine it,
let us consider a potential (\ref{eq:potential}) with $W=0$; this
amounts to setting $t_j=0$ for all $j\geq 0$. In this case, the
associated Riemann surface (\ref{polycurve}) has a single independent
period, namely $S_0$. It follows that the function $\Psi$ does not
depend on the variables $S_1\dots S_\infty$. Thus one has
$\Psi=\Psi(t_{-1},S_0)$. Returning to equation (\ref{F_10}), we now
take $W$ to be of the form (\ref{eq:potential}) and obtain: \be
\label{F_1more}
F_1(t_{-1},t_0\dots t_d,S_0\dots S_{d+\delta+1})=
\frac{s}{2}\frac{\partial F_0(t_{-1},t_0\dots t_{d}, S_0\dots
  S_{d+\delta+1})}{\partial t_{-1}}+ \Psi(t_{-1},S_0)~~.  \ee Since we
are ultimately interested in the case $t_{-1}=0$, let us set
$S_0=S_0(t_{-1})$ with $S_0(0)=0$ in equation (\ref{F_1more}). Then
the limit $t_{-1}\rightarrow 0$ gives: 
\be 
F_1(0,t_0\dots t_d,0,S_1\dots S_{d+\delta+1})= \frac{s}{2}
\left[\frac{\partial F_0(t_{-1},t_0\dots t_d, S_0\dots
S_{d+\delta+1})}{\partial t_{-1}}\right]
\Big{|}_{t_{-1}=0,S_0=0}+ \Psi(0,0)~~. 
\ee 
Therefore, the microcanonical generating function of the model with
${t_{-1}=0}$ satisfies:
\be
\label{F_1}
F_1=\frac{s}{2}\frac{\partial F_0}{\partial t_{-1}}
\Big{|}_{t_{-1}=0}+{\rm  constant}~~.
\ee 
This relation is reminiscent (though somewhat different in character) of 
similar equations satisfied by the (microcanonical) generating function of
$SO(N)$ and $Sp(N/2)$ one-matrix models \cite{Ashok:2002bi,so6}.

\section{Comparison with the field theory description}\label{sec:comp-field}

In this section we want to compute the leading terms in the effective
superpotential from our matrix model.  The strategy to do so will be
the following. We will take the superpotential of the gauge theory in
which all the fields are complex. So to start with, there is no
hermiticity constraint on $M$ and also $Q$ and $\bar Q$ are unrelated.
The gauge symmetry will be given by the complexified gauge group just
as in the discussion of the classical moduli space of the gauge
theory. After imposing an appropriate gauge fixing \`a la BRST we then
choose a suitable real section in the configuration space and perform
the path integral along this real section. This is completely
analogous to the situation in \cite{DGKV, KMT} where in a two-cut
vacuum of the one-matrix model one was forced to expand around
Hermitian and anti-Hermitian matrices respectively. This somewhat ad
hoc procedure has been justified recently in \cite{calin}.

According to the DV-conjecture the leading logarithmic terms in the
superpotential can be computed from the ``non-perturbative'' part of
the matrix-model partition function. Actually the notion of
non-perturbative should be understood here in the sense that we
compute the one-loop approximation to the partition function, taking
all couplings of terms involving the product of more than two matrices
to zero. In other words, we will expand the matrix model around the
background only up to terms that are bilinear in the fluctuating
fields.  In order to keep things simple but also capture the essential
physics we choose to expand around the vacuum with vevs given by
\begin{equation}\label{Mmatrixvevs}
  \langle M \rangle = \mathrm{diag}({\bf 0}_{N_0},
  a {\bf 1}_{N_1}, b {\bf 1}_{N_2}, -b {\bf 1}_{N_2})\,;
\end{equation}
and
\begin{equation}\label{Qmatrixvevs}
\langle Q \rangle  = \left( \begin{array}{cccc}
 \mathrm{\bf E}_s & & & \\
 & {\bf 0} & &  \\
 & & {\bf 0}& {\bf 1}_{N_2}   \\
 & & s {\bf 1}_{N_2}& {\bf 0}
\end{array} \right) ; \qquad
\langle \bar Q \rangle = \left( \begin{array}{cccc}
 -s W'(0)\mathrm{\bf E}_s & & & \\
 & {\bf 0} & &  \\
 & & {\bf 0}& -s W'(b){\bf 1}_{N_2}   \\
 & & -W'(b) {\bf 1}_{N_2}& {\bf 0}
\end{array} \right)
\end{equation}
The $N_0\times N_0$ matrix $\mathrm{\bf E}_s$ is defined as in Section
2.  The constants $a$ and $b$ are solutions to the equations $W'(a)=0$
and $W'(b)=W'(-b)$.  This vacuum is the simplest generic one in the
sense that each of the different types of vacua we found in the
analysis of the classical moduli space appears precisely once.
\subsection{Fixing the gauge \`a la BRST}
The gauge transformation are
\begin{equation}\label{gaugetrafos}
\delta M = i [ \Lambda, M]\,, \qquad \delta Q = i ( \Lambda Q + Q \Lambda^T)
\, \qquad \delta \bar Q = -i (\Lambda^T \bar Q + \bar Q \Lambda )
\end{equation}
Since we are interested only in the quadratic part of the action we
can also linearize the gauge transformations around the background.
Furthermore from now on we will decompose all matrices in blocks of
sizes $N_i\times N_j$ with $i,j \in 0\cdots 3$ and denote the
fluctuations of the matrices around the vevs with lower case letter,
i.e.  $M = \langle M \rangle + m$, $Q = \langle Q \rangle + q$ and
$\bar Q = \langle \bar Q \rangle + \bar q$. The linearized gauge
transformations then take the form
\begin{eqnarray}\label{gaugetrafoslinear}
\hat{\delta} m_{ij} &=& i \big(\langle M \rangle_{jj} - \langle M
\rangle_{ii}\big) \Lambda_{ij} \label{deltaMlin}\\ 
\hat{\delta} q_{ij} &=& 
i \big(\Lambda_{ik} \langle Q \rangle_{kj} + 
\langle Q \rangle_{ik} (\Lambda_{jk})^T \big)
\label{deltaQlin}\\
\hat{\delta} \bar q_{ij} &=& -i\big((\Lambda_{ki})^T \langle \bar Q \rangle_{kj}
+ \langle \bar Q \rangle_{ik} \Lambda_{kj} \big) \label{deltaQbarlin}
\end{eqnarray}
{}From these transformation we see that we can choose the gauge
\begin{eqnarray}\label{gaugecondm}
m_{ij}&=& 0\,;\; \mathrm{for}\; j \ne i \,,\nonumber\\ q_{23}&=&q_{00}=0\,.
\end{eqnarray}
The gauge fixing is now implemented in a standard fashion by promoting
the gauge transformations to BRST transformations, i.e.\ replacing the
gauge parameters by ghosts. We also need to introduce antighosts $\bar
C$ and Lagrange multiplier fields $B$. The gauge fixing Lagrangian is
given by
\begin{equation}
S_{gf} = \mathbf{s}\, \Tr \left( \sum_{i\ne j} (\bar C_{ij} m_{ij}) + \bar
C_{q} q_{23} + \bar C_{00} q_{00} \right)\,.
\end{equation}
The BRST transformations for the antighosts and Lagrange multipliers
are
\begin{eqnarray}
\mathbf{s}\, \bar C_{ij} = B_{ij}\,, & \mathbf{s}\, \bar B_{ij} = 0\,,
\nonumber \\ \mathbf{s}\, \bar C_{q} = B_{q}\,, & \mathbf{s}\, \bar B_{q} =
0\,, \nonumber \\ \mathbf{s}\, \bar C_{00} = B_{00}\,, & \mathbf{s}\, \bar
B_{00} = 0\,.
\end{eqnarray}
Explicitly the gauge fixing action is
\begin{eqnarray}\label{gaugefixingaction}
S_{gf} &=& \Tr \Big[ \sum_{i \ne j} B_{ij}m_{ij} + i (\langle M \rangle_{ii} -
\langle M \rangle_{jj}) \bar C_{ij} C_{ij} + B_q q_{23} -i \bar C_q (C_{22} +
C_{33}^T) + \nonumber\\ & & B_{00} q_{00} -i \bar C_{00} (C_{00} \mathrm{\bf
E}_s +\mathrm{\bf E}_s C_{00})\Big].
\end{eqnarray}
The the residual gauge group is
\begin{equation}
G_s = U(N_1)\times U(N_2) \times \left\{ \begin{array}{cc} SO(N_0) & s=+1 \\
Sp(\frac{N_0}{2}) & s=-1
\end{array}\right.\,.
\end{equation}
Notice that the $U(N_2)$ group is diagonally embedded in $U(N_2)\times
U(N_2)$ such that the rank of the original $U(N)$ gauge group is
$N=N_1 + 2 N_2 + N_0$.  According to this structure of the gauge group
only the combination $C_{22}+C_{33}^T$ appears in the gauge fixed
action.  Similarly only the ghosts components of $C_{00}$ that do not
correspond to $SO/Sp$ residual gauge transformations propagate.

\subsection{Gaussian approximation of partition function and microcanonical
generating function}
Expanding the action around the background up to terms that are
bilinear in the fields gives
\begin{eqnarray}
S^{(2)}& =& N\left(N_1 W(a) + N_2(W(b)+W(-b)) + N_0 W(0) + \frac 1 2
\sum_{i=0}^3 W''(\langle M \rangle_{ii}) \Tr( m_{ii}^2 )\right) + \nonumber \\
& & \sum_{i > j} (\langle M \rangle_{ii}+\langle M \rangle_{jj}) \Tr(\bar
q_{ij} q_{ji}) + \sum_{i=0}^3 \langle M \rangle_{ii} \Tr(\bar q_{ii} q_{ii}) +
\nonumber \\ & & +\Tr ( \bar q_{32} (m_{22} + m_{33}^T) ) + \frac 1 2 \Tr
(\bar q_{00}(m_{00} \mathrm{{\bf E}}_s + \mathrm{{\bf E}}_s m_{00}^T) )\,.
\end{eqnarray}
In this expression we used already that $q_{23}=0$, $q_{00}=0$ and
$m_{ij} =0$ for $i\neq j$.  Notice that this gauge fixing is also
important for another reason.  Even without our gauge choice the
blocks $q_{23}$, $q_{00}$ do not contribute to $S^{(2)}$. This is
because $\langle M \rangle_{22} = b = - \langle M \rangle_{22}$ and
$\langle M \rangle_{00}=0$.  Fixing the gauge in the way we did
eliminates this problem.  Moreover, the fields $\bar q_{23}$ and $\bar
q_{00}$ act now as Lagrange multipliers implementing the constraints
\begin{equation}\label{Mconstraints}
m_{22} = - m_{33}^T \qquad \mathrm{and}\qquad m_{00}\mathrm{\bf E}_s +
\mathrm{\bf E}_s m_{00}^T=0\,.
\end{equation}
This restricts the fluctuations of $M$ precisely in such a way as to
account for a field in the adjoint representation of the diagonally
embedded $U(N_2)$ gauge group and another field in the adjoint
representation of $SO(N_0)$ or $Sp(\frac{N_0}{2})$ respectively.

In the definition of the matrix model in section 2 we had to implement
also a reality constraint of the form $M-M^\dagger=2 i \eps {\bf 1}$
and $\bar Q = -i Q^\dagger$.  However, for the present calculation it
is more convenient to choose different sections to ensure that all
exponentials are decaying.  For instance if $a$ sits at a maximum of
$W(x)$ we will choose $m_{11}$ anti-Hermitian.  Similarly, we choose
$\bar q_{ij} = q_{ij}^\dagger$ or $\bar q_{ij}= - q_{ij}^\dagger$
depending on the sign of the coefficient of $\bar q_{ij} q_{ij}$ in
$S^{(2)}$. For the ghosts we choose either $\Re (C_{ij}) = \Re (\bar
C_{ij}) =0 $ or $\Im (C_{ij}) = \Im (\bar C_{ij}) =0 $.

In addition we can not relate all blocks of $q$ to blocks of $\bar q$
by a conjugation. Demanding for example $\bar q_{32}=q_{23}^\dagger$
would automatically set $\bar q_{32}$ to zero because of the gauge
fixing.  We do need these fields however in order to implement the
constraints (\ref{Mconstraints}) on the fluctuations of $M$.
Therefore we will demand ${\bar q}_{32}$ imaginary and ${\bar q}_{00}$ 
imaginary
symmetric or antisymmetric depending on $s$. After integrating out the
Lagrange multipliers $B_{ij}$, $B_q$, $B_{00}$, $\bar q_{32}$ and
$\bar q_{00}$ the partition function of the gauge fixed matrix model
is given by
\begin{equation}
Z = \frac{1}{\mathrm{vol}(G_s)} \int \prod_{i\in\{0,1,2\}}dm_{ii} \prod_{ij \ne
{\tiny \left\{\begin{array}{c}(23)\\(00)\end{array}\right.}} dq_{ij}
\;d C \;d\bar C \;e^{(S^{(2)}+S_{gf})}
\end{equation}
Performing the Gaussian integrations gives
\begin{equation}
Z = \frac{1}{\mathrm{vol}(G_s)} Z_{M}\, Z_{ghosts}\, Z_{matter}\,,
\end{equation}
where
\begin{eqnarray}
Z_{M} &=& \Big( \frac{2 \pi}{W''(a)} \Big)^{\frac{N_1^2}{2}}
\Big( \frac{2 \pi}{W''(b)+W''(-b)} \Big)^{\frac{N_2^2}{2}}
\Big( \frac{2 \pi}{W''(0)} \Big)^{\frac{N_0}{2}(N_0-s)}.\nonumber\\
Z_{ghosts} &=& (a^2-b^2)^{2N_1 N_2} a^{2 N_1 N_2} (2b)^{2 N_2^2} b^{4N_2 N_0}\nonumber \\
Z_{matter} &=&  \Big( \frac{ \pi}{a^2-b^2} \Big)^{N_1 N_2} \Big( \frac{ \pi}{a} \Big)^{N_1 N_2+\frac{N_1}{2}(N_1+s)}
\Big( \frac{ \pi}{b} \Big)^{2N_2 N_0+N_2(N_2+s)}\,.
\end{eqnarray}
The microcanonical generating function is defined through $N^2 F = -
\log(Z)$. In order to compute it we need the logarithms of the volume
of the residual gauge group. These are given by
\begin{eqnarray}
\log(\mathrm{vol}(U(N))) &=& -\frac{N^2}{2}\left( \log(N) -\frac 3 2 - \log(2\pi) \right)+O(N^0)\,,\nonumber\\
\log(\mathrm{vol}(H_s(N))) &=& -\frac{N^2}{4} \left( \log(N) - \frac 3 2 - \log(2\pi) \right) +\nonumber\\
& & s \frac{N}{4} \left(\log(N)-1-\log(4)-\log(2\pi)\right) + O(N^0)\,,
\end{eqnarray}
where $H_s$ is $SO(N)$ for $s=1$ and $Sp(\frac{N}{2})$ for $s=-1$.
According the the DV conjecture we introduce now the filling fractions
$S_i = \frac{N_i}{N}$ and define the masses $m(a) = W''(a)$, $m(b) =
(W''(b)+W''(-b))$ and $m(0)= W''(0)$.  The $1/N$ expansion of the
microcanonical generating function we denote by
\begin{equation}F= N^2 \left(F^0 + \frac{1}{N} F^1 + O(1/N^2) \right)\,.\end{equation}
The complete microcanonical generating function is
\begin{equation}
F=F_{M} + F_{ghosts} + F_{matter}
\end{equation}
where
\begin{eqnarray}
F_{M} &=& -\frac{S_1^2}{2} \log\Big( \frac{S_1}{e^{\frac 3 2} m(a)}\Big) -
\frac{S^2_2}{2} \log\Big( \frac{S_2}{e^{\frac 3 2} m(b)}\Big) -  \frac{S_0^2}{4} \log\Big( \frac{S_0}{e^{\frac 3 2} m(a)}\Big) +\nonumber \\
& & +\frac{s}{N} \frac{S_0}{4} \log\Big( \frac{4 S_0}{e m(0)}\Big)  \,,\nonumber\\
F_{matter} &=& -S_1 S_2 \log\Big( \frac{\pi^2}{a^2-b^2}\Big) -(S_1 S_0+S_2^2/2-\frac{s}{2N} S_1)
\log\Big( \frac{\pi}{a}\Big)- \\
& & - (2S_2 S_0 + S_2^2 +s S_2) \log( \frac{\pi}{b}\Big) \,,\nonumber\\
F_{ghosts} &=& -2 S_1 S_2 \log(a^2-b^2) -2 S_1 S_0 \log(a) - 2 S_2^2 \log(2b) -4 S_2 S_0 \log(b)\,. \nonumber
\end{eqnarray}

\subsection{The superpotential in the Gaussian approximation}
The effective superpotential of the gauge theory can be computed from
these expressions as
\begin{equation}\label{Weff}
W_{eff} = \sum_{i\in\{1,2,0\}} N_i \frac{\partial F^0}{\partial S_i} + 4 F^1 + \alpha_i S_i\,,
\end{equation}
The subleading $1/N$ term in the microcanonical generating function
stems form graphs with topology $\R\P^2$. These are present in our
theory because of the symmetric/antisymmetric $Q$ matrices. It has
been argued in \cite{Ita:2002kx, Ashok:2002bi, so6} that the
$\R\P^2$ contribution to $F$ enters the superpotential with a an
additional factor $4$. This argument has originally been derived in
the context of theories with orthogonal and symplectic gauge groups.
However, it really only depends on the topology of the Feynman graphs
and therefore should go over to our case without essential changes. In
any case we will see shortly that (\ref{Weff}) gives the correct
Veneziano-Yankielowicz form of the superpotential. The final
ingredient is $\alpha$.  It is given by the coupling of the original
$U(N)$ gauge theory
\begin{equation}
\alpha = (N-2s) \log\Big(\frac{\Lambda_{\mathrm{high}}}{\mu}\Big)\,.
\end{equation}
The factor $N-2s$ is the one-loop beta function coefficient of the
original gauge theory.  The scale $\mu$ is the scale where the
potential $W(\phi)$ becomes relevant. For notational simplicity we
will set this scale to one. Alternatively we could have introduced a
scale in the matrix integral such that the $S_i$ obtain the correct
dimension three as is appropriate for the interpretation as gaugino
condensate.  The $\alpha_i$ are then given by $\alpha_{0,1} = \alpha$
and $\alpha_2=2\alpha$, where the last factor two stems from the
diagonal embedding of the gauge group $U(N_2)$.  Collecting now all
the terms we find the effective low energy superpotential
\begin{eqnarray}\label{Weff-2}
W_{eff} &=& S_1 \log \Big( \frac{\Lambda_{\mathrm{low,1}}^{3N_1} } {S_1^{N_1}}
\Big) + S_2 \log \Big( \frac{ \Lambda_{\mathrm{low,2}}^{3N_2} } {S_2^{N_2}}
\Big) + \frac{S_0}{2} \log \Big( \frac{\Lambda_{\mathrm{low,0}}^{3(N_0-2s)} }
{S_0^{N_0-2s}} \Big) \nonumber \\ & & + S_1 \log\left(\frac{e^{N_1}}{
\pi^{N+2s}} \right) + S_2 \log\left(\frac{e^{N_2}}{\pi^{2(N-2s)-2N_2}}\right)
+ \frac{S_0}{2} \log\left( \frac{4^{2s}e^{N_0-2s}}{ \pi^{2N_1+4N_2}} \right)
\,.
\end{eqnarray}
The relation between the low- and high energy scales is
\begin{eqnarray}
\Lambda_{\mathrm{low,1}}^{3N_1} &=& m(a)^{N_1}\, a^{N_1+2s}\, (a^2-b^2)^{-N_2}
\,a^{-N_0}\, \Lambda_\mathrm{high}^{N-2s} \,,\nonumber\\
\Lambda_{\mathrm{low,2}}^{3N_2} &=& m(b)^{N_2}\, (a^2-b^2)^{-N_1}\,
b^{2N_2+4s-2N_0}\, (2b)^{-4N_2}\,
\Lambda_\mathrm{high}^{2(N-2s)}\,,\nonumber\\
\Lambda_{\mathrm{low,0}}^{\frac{3}{2}(N_0-2s)} &=& m(0)^{\frac{N_0}{2}-s}\,
a^{-N_1}\, b^{-2N_2}\, \Lambda_\mathrm{high}^{N-2s}\,.
\end{eqnarray} 
Thus the effective superpotential coincides with the one obtained in
section 2 up to the linear terms in the $S_i$. Of course these linear
terms can not be determined by naive threshold matching.

\bigskip

\acknowledgments{ We thank Stefan Theisen for numerous helpful discussions.
We also thank  Gernot~Akemann and Gregory~Moore for email correspondence.
This work was supported by DFG project KL1070/2-1. }

\appendix

\section{Comparison with the $O(n)$ and $A_2$ matrix models}

\subsection{Comparison with the $O(n)$ model}

Let us compare our models with symmetric and antisymmetric matter with
the $O(n)$ model analyzed in \cite{EK}. The difference is the $(M,Q)$
interaction term, which reads in the $O(n)$ model
\begin{equation}
  \sum_{i=1}^n \tr Q_i^\dag M Q_i,
\end{equation} 
where $Q_i^\dag =Q_i $ are Hermitian.  After diagonalizing $M$ in the
relevant case for comparison\footnote{For a special cubic $V(x)$ this
  corresponds to Ising model on a random lattice
  \cite{Eynard:1992cn}.}  $n=1$, the Gaussian integration over the
real $Q_{ii}$ leads in our conventions to $s=0$. To summarize one has
$s=1$ for $Q_{ii}$ complex, $s=0$ for $Q_{ii}$ real and $s=-1$ for
$Q_{ii}=0$ and this affects only the definition of $U(x)$
(\ref{eq:U-def}) while the rest of the derivation of the loop
equations remains the same. The quadratic loop equation (\ref{eq:l2})
agrees in the large $N$-limit with the one in \cite{EK}. A cubic loop
equation does not appear in \cite{EK} explicitly. The second order
loop equations generalize very directly to general $n$. The third
order loop equation should become an $n+1$ order loop equation. {}From
the gauge theory point of view the $n=1$ model corresponds to adding
matter in the adjoint representation, a spectrum which emerges e.g.
from an ${\cal N}=4$ theory.  Adding more matter $n>1$ leads to non
asymptotically free cases.

\subsection{Comparison with the $A_2$ quiver theory}

Let us next compare the situation to the $ADE$ quiver theories, in
particular to $A_2$, which is the double cover of the models discussed
so far.  The $ADE$ matrix models where discussed in \cite{Kostov_ade,
  Kostov_stat, Kharchev} and further considered in the context of the
DV conjecture in \cite{DV2, Hofman, Seki}.  The action of the $ADE$
quiver matrix models is: \be S_{\mathrm{ADE}}=\sum_{i=1}^r \tr
W_{(i)}(M^{(i)}) + \sum_{i,j=1}^r s_{ij}\tr Q_{(ij)} M^{(j)} Q_{(ji)}\ 
.
\label{quiveraction}\ee 
Here one has an $N^{(i)} \times N^{(i)}$ matrix $M^{(i)}$ for every
node in the Dynkin diagram of the $ADE$ Lie algebra and the $Q_{(ij)}$
are $N^{(i)}\times N^{(j)}$ matrices transforming in the
$(N^{(i)},\bar N^{(j)})$ representation and fulfilling
$Q^\dag_{(ij)}=Q_{(ji)}$. For a given ordering of the nodes
$s_{ij}=-s_{ji}=1$ if node $i$ and node $j$ are linked in the diagram
and $s_{ij}=0$ otherwise.  After diagonalizing all $M^{(i)}$ the
Gaussians in the $Q$-fields appear with different signs.  Let us
assume $s_{ij}=1$ and $Q_{ji}=i Q_{ij}^\dagger$.  Then we have to
shift the $M^{(i)}$ eigenvalues into the upper half-plane whereas the
$M^{(j)}$ eigenvalues have to be shifted into the lower half-plane !
For $s_{ij}=-1$ we can still shift the $M^{(i)}$ eigenvalues into the
upper half-plane but we then have to demand $Q_{ji}=-i
Q_{ij}^\dagger$.  After choosing these real slices through the matrix
configuration space the Gaussian integrations over all $Q$-fields can
be performed.  \be N^2 S_{\mathrm{ADE}}=N\sum_{i=1}^r
\sum_{k=1}^{N^{(i)}} W_{(i)} (\lambda_k^{(i)})- \sum_{i=1}^r
\sum_{k\neq l}\ln(\lambda_k^{(i)}-\lambda_l^{(i)})+\sum_{i<j}^r
\sum_{k,l}\ln(\lambda_k^{(i)}- \lambda_l^{(j)})^{|s_{ij}|}\ .
\label{intquiveraction}\ee
The orientifold projections are obtained by
$M^{(1)}=-\left(M^{(2)}\right)^T$ and $\bar Q_{12}^T=s Q_{21}$ and
lead to the symmetric or antisymmetric matter. Identifying
$Q_{12}=Q_{21}$ leads to the $n=1$ $O(n)$ model. The action on $M$
identifies for all cases $\lambda^{(1)}_k=-\lambda^{(2)}_k$, while the
projection of the $Q$ changes the result of the Gaussian integration
starting from (\ref{quiveraction}).  Let us sketch the derivation of
the exact loop equations for $S:=S_{A_2}$ in the following. The e.o.m.
for the $\lambda^{(1)}_k$ eigenvalues read: 
\be 
N{\partial
  S_{\mathrm{ADE}} \over \partial \lambda^{(1)}_k}=W_{(1)}'(\lambda^{(1)}_k) 
-{1\over
  N} \sum_{l(\neq k)}^{N_1} {2 \over \lambda^{(1)}_k -
  \lambda^{(1)}_l} +{1\over N} \sum_{l}^{N_2} {1 \over \lambda^{(1)}_k
  - \lambda^{(2)}_l}, 
\ee 
with an analogous expression for the $\lambda^{(2)}_k$.

Let us define the resolvents for the two matrices as
$\omega_i(z)={1\over N} \sum_{k}^{N_i} {1\over z-\lambda^{(i)}_k}$,
$i=1,2$ and choose $\Psi_k^{(i)}={1\over z- \lambda^{(i)}_k}$, $i=1,2$
in (\ref{eq:Ward}) to obtain two Ward identities, which can be
simplified with analogous equations to (\ref{eq:2lam-ident}). Adding
these two Ward identities one gets the quadratic loop equation: \be
\langle \omega_1(z)^2 - \omega_1(z)\omega_2(z) + \omega_2(z)^2 -
W'_1(z)\omega_1(z) - W'_2(z) \omega_2(z)+f_1(z) + f_2 (z)\rangle=0 \ .
\label{quadraticloopA2}
\ee Here we defined \be f_i(z)={1\over N} \sum_{k=1}^{N_i} {W'_i(z) -
  W'_i (\lambda^{(i)}_k) \over z- \lambda^{(i)}_k}, \quad i=1,2\ .
\ee

The derivation of the cubic loop equations is likewise very similar to
the discussion in the orientifolded model. One uses
$\Psi^{(1)}_k={1\over N} \sum_{m=1}^{N_2} {1\over \lambda^{(1)}_k -
  \lambda^{(2)}_m} {1\over z- \lambda^{(1)}_k}$ in (\ref{eq:Ward}) and
simplifies the Ward identity with (\ref{cubfrac}).  Subtracting the
same equation, but with indices $(1)\, \leftrightarrow\, (2)$
exchanged, yields the cubic loop equation: \bea
\biggl\langle\omega_1(z)\omega_2(z)^2- \omega_1(z)^2\omega_2(z)
+W'_1(z) (\omega_1(z)^2+ f_1(z) -W'_1(z) \omega_1(z))- \nonumber \\ \ 
W'_2(z) (\omega_2(z)^2+ f_2(z) -W'_2(z) \omega_2(z)) + g_2(z) - g_1(z)
\biggr\rangle \ =0,
\label{cubicloopA2}
\eea where \be g_i(z)={1\over N^2} \sum_{k=1}^{N_1} \sum_{m=1}^{N_2}
{W'_i(z)-W'_i(\lambda^{(i)}_k)\over (\lambda^{(i)}_k-\lambda^{(j\neq
    i)}_m)(z-\lambda_k^{(i)})}\ , \ee
Note that the cubic loop equation for the quiver model does not depend
explicitly on $\frac{1}{N}$.  In contrast the cubic loop equation for
the $O(1)$ model does contain a subleading $\frac{1}{N^2}$ term and
the orientifold model contains even terms at order $\frac{1}{N}$. This
is consistent with the fact that $\frac{1}{N}$ terms come from
non-orientable diagrams and these are absent of course in the $A_2$
and $O(1)$ model.  According to the DV conjecture, terms of order
$\frac{1}{N^2}$ are related to gravitational corrections.  The
effective pure field theoretical superpotential of the orbifold model
can therefore be obtained by taking the effective superpotential of
the $A_2$ model and simply identifying the relevant variables, e.g.
$S_i^{(1)}=S_i^{(2)}$.  However, in the terms related to the
gravitational couplings this procedure would not give a correct
result, since then one has to take into account the additional
$\frac{1}{N^2}$ terms in the cubic loop equation!  It would be
interesting to investigate this issue further and to see if these
extra terms can be related to the presence of twisted sectors in the
dual string model.

\subsubsection{The large $N$ Riemann surface for the $A_2$-quiver}\label{sec:large-N-A2}

Here we will take the large $N$ limit of the loop equations for the
$A_2$ quiver. We will see that the Riemann surface governing the $A_2$
model is not hyperelliptic but rather is a triple cover of the
$z$-plane, one of whose branches gives $\om_{i,0}(z)$.  The occurrence
of Riemann surfaces with a more complicated sheet structure for multi
matrix models is not unexpected \cite{Kostov_cft, DV2}.  Our
derivation from the exact loop equations will provide a proof for the
curves of $A_2$ quiver and the orbifolds thereof.

The leading terms will be extracted from the loop equations
(\ref{quadraticloopA2},\ref{cubicloopA2}).  Using the
variables\footnote{In order to keep the notation simple, we use here
  $\om_i$ for $\om_{i,0}$.}  \bea
\omega_1(z)&=&u_1(z) - t_1(z)\ \ , \quad t_1(z)=-{1\over 3} \left(2 W'_1(z) + W'_2(z)\right) \nonumber \\
\omega_2(z)&=&-u_2(z) + t_2(z)\ \ , \quad t_2(z)={1\over 3}
\left(W'_1(z) + 2 W'_2(z)\right)
\label{shift}
\eea one obtains the from the large $N$-loop equations the affine
complex algebraic curve \bea
u_1(z)^2 + u_1(z) u_2(z)+u_2(z)^2&=&p(z)\nonumber\\\
u_1(z) u_2(z)^2 + u_1(z)^2 u_2(z)&=&-q(z)
\label{cis}
\eea where \bea
p(z)&=&t_1^2+ t_1 t_2+ t_2^2 - f_1-f_2\nonumber\\
q(z)&=&-t_1^2 t_2- t_1 t_2^2 + f_1 t_2 + f_2 t_1 - g_1 + g_2\ .
\label{pqdef}
\eea Now (\ref{cis}) are the Viete relations for the cubic \bea
\prod_{i=0}^2 (u(z)-u_i)&=& u^3-p(z) u - q(z) \nonumber\\
&=&\prod_{i=0}^2 (u-t_i(z))+u(f_1+f_2)+g_1-g_2-f_2 t_1 - f_1 t_2 =0\ ,
\label{cubicA2} \eea where $u_0=-u_1-u_2$ and $t_0=-t_1-t_2$.  As we
have shown that $f_i$ and $g_i$ are polynomials in $z$, \erf{cubicA2}
is an algebraic curve, and can be viewed as a triple cover of the $z$
plane \cite{DV2}. The perturbations encoded in $f_i$ and $g_i$ can be
explicitly related to the eigenvalue densities of the two matrix
model.
If $d_1+1, d_2+1$ are the degrees of the potentials $W_1,W_2$, the
degrees of $f_i,g_i$ are $d_i-1$ respectively. The perturbation
parameters introduced by $g_i$ are $\#(g)={\rm max} (d_1,d_2)$ and
$\#(f)=d_1+d_1$ for the $f_i$. In particular for $d_1=d_2=d$ relevant
for the modding below we get $3 d$, which is is compatible with number
of branches calculated in \cite{Cachazo:2001gh}: for equal degree
$d+1$ of the $W_i$ it is $d R_+$, where $R_+$ are the number of
positive roots for the quiver ADE group.  It has also been checked
\cite{Cachazo:2001gh} that this number coincides with the number of
independent $S^3$ in $H^3(M)$, where M is the local Calabi-Yau
threefold after the large $N$ transition.

Using the $\Z_2$ identification $\lambda^{(1)}=-\lambda^{(2)}$ on gets
\bea \om_1(z)&=&\om(z), \ \ \ \om_2(z)=-\om(-z), \\ \nonumber
W'_1(z)&=&W'(z), \ \ \ W'_2(z)=-W'(-z) \label{Z2identifications} \\ 
\nonumber \eea etc. Since it was shown in section \ref{sec:constr}
that all log dependence can be absorbed into $V(z)$ we can in general
replace $W'(z)$ by $V'(z)$ and recover the large $N$ limit of the two
loop equations with the $\Z_2$ orientifold identification
\erf{eq:genus0}.

\end{document}